%% file: SUSY-2016-22-PAPER.tex
\newcommand*{\ATLASLATEXPATH}{latex/}
\pdfoutput=1
\documentclass[PAPER,texlive=2016, UKenglish,cernpreprint,]{\ATLASLATEXPATH atlasdoc}
\usepackage[subfigure=true, backend=biber]{\ATLASLATEXPATH atlaspackage}
\usepackage{\ATLASLATEXPATH atlasbiblatex}
\usepackage{\ATLASLATEXPATH atlasphysics}
\usepackage{\ATLASLATEXPATH atlasbsm}
\usepackage{\ATLASLATEXPATH atlasjetetmiss}
\usepackage{\ATLASLATEXPATH atlasmisc}
\graphicspath{{logos/}{figures/}}

\usepackage{SUSY-2016-22-PAPER-defs}
\usepackage{a4wide}
\usepackage[T1]{fontenc}
\usepackage{color}
\usepackage{siunitx}
\usepackage{booktabs}
\usepackage{dcolumn}
\usepackage{amssymb}

\usepackage{multirow}

\input{SUSY-2016-22-PAPER-metadata}

\hypersetup{pdftitle={ATLAS document},pdfauthor={The ATLAS Collaboration}}

\DeclareOldFontCommand{\rm}{\normalfont\rmfamily}{\mathrm}
\DeclareOldFontCommand{\tt}{\normalfont\ttfamily}{\mathtt}

\begin{document}

\maketitle

\section{Introduction}
\label{sec:intro}

\begin{figure}[!ht]
    \centering
  \subfigure[gluino direct decay]{
   \includegraphics[width=0.4\textwidth]{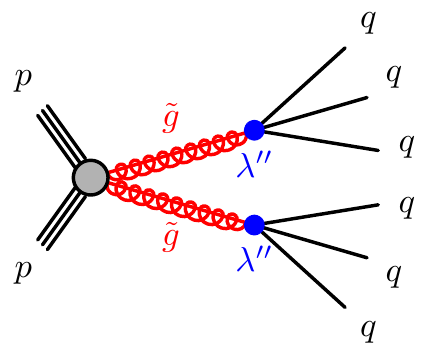}
    \label{fig:Q6}}
  \subfigure[gluino cascade decay]{
   \includegraphics[width=0.4\textwidth]{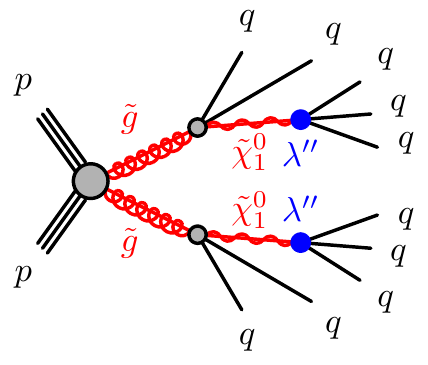}
    \label{fig:Q10}}
      \caption{Diagrams for the benchmark processes considered for this analysis. The black lines represent Standard Model particles, the red lines represent SUSY partners, the grey shaded circles represent effective vertices that include off-shell propagators (e.g. heavy squarks coupling to a \ninoone neutralino and a quark), and the blue solid circles represent effective RPV vertices allowed by the baryon-number-violating \lampp\ couplings with off-shell propagators (e.g. heavy squarks coupling to two quarks). Quark and antiquark are not distinguished in the diagrams. (For interpretation of the colours in the figure(s), the reader is referred to the web version of this article.)}
    \label{fig:intro:diagram}  
\end{figure}

Supersymmetry (\SUSY)~\cite{Golfand:1971iw,Volkov:1973ix,Wess:1974tw,Wess:1974jb,Ferrara:1974pu,Salam:1974ig}
is a theoretical extension of the Standard Model (SM) which fundamentally relates fermions and bosons. It is an alluring theoretical possibility given its potential to solve the hierarchy problem\cite{Sakai:1981gr,Dimopoulos:1981yj,Ibanez:1981yh,Dimopoulos:1981zb}.
This Letter presents a search for supersymmetric gluino pair production with subsequent R-parity-violating (RPV)~\cite{Farrar:1978xj,Hall:1983id,Ross:1984yg,Barger:1989rk,Dreiner:1997uz,Barbier:2004ez} decays into quarks in events with many jets 
using \lumia~of $p$--$p$ collision data at $\sqrt{s}$ = 13 TeV collected by the ATLAS detector in 2015 and 2016. 
In the minimal supersymmetric extension of the Standard Model, the RPV component of a generic superpotential can be written as~\cite{Dreiner:1997uz, Allanach:2003eb}:
  \begin{eqnarray} 
      W_\mathrm{RPV} &= \frac{1}{2} \lamijk L_{i}L_{j}\Ebar_{k} + \lampijk L_{i}Q_{j}\Dbar_{k} + \frac{1}{2}\lamppijk\Ubar_{i}\Dbar_{j}\Dbar_{k} + \kappa_iL_{i}H_{2},
     \label{eq:rpvandudd:wrpv}
   \end{eqnarray} 
\noindent where $i,j,k=1,2,3$ are generation indices. The generation indices are omitted in the discussions that follow if the statement being made is not specific to any generation. The first three terms in \equref{rpvandudd:wrpv} are often referred to as the trilinear couplings, whereas the last term is referred to as bilinear. The $L_i$ and $Q_i$ represent the lepton and quark $SU(2)_{\rm L}$ doublet superfields, whereas $H_2$ represents the Higgs superfield. The $\Ebar_j$, $\Dbar_j$, and $\Ubar_j$ are the charged lepton, down-type quark, and up-type quark $SU(2)_{\rm L}$ singlet superfields, respectively. The couplings for each term are given by \lam, \lamp, and \lampp, while $\kappa$ is a mass parameter. 
In the benchmark models considered in this search, the couplings of \lam and \lamp are set to zero and only the baryon-number-violating coupling $\lamppijk$ is non-zero.
Because of the structure of \equref{rpvandudd:wrpv}, scenarios in which only $\lamppijk \neq 0$ are often referred to as UDD scenarios.
The diagrams shown in Figure~\ref{fig:intro:diagram} represent the benchmark processes used in the optimization and design of the search presented in this Letter. 
In the gluino direct decay model (Figure~\ref{fig:Q6}), the gluino directly decays into three quarks via the RPV UDD coupling \lampp, leading to six quarks at tree level in the final state of gluino pair production.
In the gluino cascade decay model (Figure~\ref{fig:Q10}), the gluino decays into two quarks and a neutralino, which, in turn, decays into three quarks via the RPV UDD coupling \lampp, resulting in ten quarks at tree level in the final state of gluino pair production. 
Events produced in these processes typically have a high multiplicity of reconstructed jets. 
In signal models considered in this search, 
the production of the gluino pair is assumed to be independent of the value of \lampp.
 Decay branching ratios of all possible \lampp flavor combinations given by the structure of~\equref{rpvandudd:wrpv} are assumed to be equal, and decays of the gluino and neutralino are implemented as prompt decays via modifying the decay widths of gluinos and neutralinos.
In this configuration, a significant portion of signal events contain at least one bottom or top quark.
Other models of the RPV UDD scenario, such as the Minimal Flavour Violation model~\cite{Nikolidakis:2007fc,Csaki:2011ge}, 
predict that the gluino decays preferentially into final states with third-generation quarks.
These theoretical arguments motivate the introduction of $b$-tagging requirements into the search.

This analysis is an update to previous ATLAS searches for signals arising from RPV UDD scenarios~\cite{SUSY-2012-16,SUSY-2013-07} performed with data taken at $\sqrt{s}$ = 8 TeV.
The search strategy closely follows the one implemented
in Ref.~\cite{SUSY-2013-07},
which excludes a gluino with mass up to 917 GeV in the \sixQ,
and a gluino with mass up to 1000 GeV for a neutralino mass of 500 GeV in the \tenQ.
Two other publications~\cite{SUSY-2016-11,SUSY-2016-14} from the ATLAS Collaboration reported on the searches for signals from a different \tenQ\ where the quarks/antiquarks from the gluino decay are top quark--anti-quark pairs and the quarks from the neutralino decays are $u$, $d$ or $s$ quarks. These searches probed events with at least one electron or muon. The most stringent lower limit on the gluino mass, from Ref.~\cite{SUSY-2016-11}, is 2100 GeV for a neutralino mass of 1000 GeV. 
In a recent publication~\cite{CMS-SUS-16-040}, the CMS Collaboration set a lower limit of 1610 GeV on the gluino mass in an RPV UDD scenario where the gluino exclusively decays into a final state of a top quark, a bottom quark and a strange quark, using $\sqrt{s} = 13$ TeV $pp$ collision data.

\section{ATLAS detector}
\label{sec:detector}

The ATLAS detector~\cite{PERF-2007-01} covers almost the whole solid angle around the collision point with layers of
tracking detectors, calorimeters and muon chambers. 
The inner detector, immersed in a magnetic field provided by a solenoid,
 has full coverage in  $\phi$ and covers the pseudorapidity range $|\eta|<2.5$.\footnote{ATLAS uses
a right-handed coordinate system with its origin at the nominal
interaction point in the centre of the detector and the $z$-axis along
the beam direction.
The $x$-axis points toward the centre of the LHC ring, and the $y$-axis
points upward.
Cylindrical coordinates $(r,\phi)$ are used in the transverse
plane, $\phi$ being the azimuthal angle around the beam pipe.
The pseudorapidity $\eta$ is defined in terms of the polar angle
$\theta$ by $\eta \equiv -\ln[\tan(\theta/2)]$.}
It consists of a silicon pixel detector, a silicon microstrip detector and
a transition radiation straw-tube tracker. 
The innermost pixel layer, the insertable B-layer, was added
between Run-1 and Run-2 of the LHC,  at a radius of 33 mm around a new, thinner, beam pipe~\cite{ATLAS-TDR-19}.
In the pseudorapidity region $|\eta| < 3.2$, high granularity
lead/liquid-argon  (LAr) electromagnetic (EM) sampling calorimeters are
used.
A steel/scintillator tile calorimeter provides hadronic calorimetry coverage
over $|\eta| < 1.7$.
The end-cap and forward regions, spanning $1.5 < |\eta| < 4.9$, are
instrumented with LAr calorimetry for both the EM and hadronic
measurements.
The muon spectrometer surrounds these calorimeters, and comprises
 a system of precision tracking chambers and fast-response detectors for triggering, with three large toroidal magnets,
each consisting of eight coils, providing the magnetic field for the muon detectors.
 A two-level trigger system is used to select events~\cite{TRIG-2016-01}. The first-level trigger is implemented in hardware
and uses a subset of the detector information. This is followed by the software-based high-level trigger, 
reducing the event rate to about 1 kHz.

\section{Simulation samples}
\label{sec:simulation}

Signal samples were produced covering a wide range of gluino and neutralino masses.
In the \sixQ,
the gluino mass (\mg) was varied from 900 GeV to 1800 GeV. In the
case of the cascade decays, for each gluino mass (1000 GeV
to 2100 GeV), separate samples were generated with multiple
neutralino masses (\mx) ranging from 50 GeV to 1.65 TeV.
In each case, \mx~<~\mg.
In the \tenQ, the two quarks produced from the gluino decay were restricted to be first or second generation quarks.
All three generations of quarks were allowed to be in the final state of the lightest supersymmetric particle decay.
Signal samples were generated at leading-order (LO) accuracy with up to two additional partons using 
the {\textsc MadGraph5\_aMC@NLO} v2.3.3 event generator~\cite{Alwall:2014hca} interfaced with \Pythia 8.186~\cite{Sjostrand:2014zea} for the parton shower, fragmentation and underlying event.
The A14 set of tuned parameters~\cite{ATL-PHYS-PUB-2012-003} was used together with the NNPDF2.3LO parton distribution function
(PDF) set~\cite{Ball:2012cx}. The EvtGen v1.2.0 program was used to describe the properties of the $b$- and $c$-hadron
decays in the signal samples. 
The signal production cross sections were calculated at next-to-leading order (NLO) in the strong coupling constant, adding
the resummation of soft gluon emission at next-to-leading-logarithm accuracy (NLO+NLL)~\cite{Beenakker:1996ch,Kulesza:2008jb,Kulesza:2009kq,Beenakker:2009ha,Beenakker:2011fu}.
The nominal cross section and its uncertainty were taken from Ref.~\cite{Borschensky:2014cia}.
Cross sections were evaluated assuming masses of 450 TeV for the light-flavour squarks
in the case of gluino pair production. 
In the simulation, the total widths of gluinos and neutralinos were set to be 1 GeV, effectively making their decays prompt. 

While a data-driven method was used to estimate the background,
simulated events were used to establish, test and validate the methodology of the analysis.
Multijet events constitute the dominant background in the search region, with small contributions from top-quark pair production ($\ttbar$).
Contributions from $\gamma$ + jets, \Wboson + jets,~\Zboson + jets, single-top-quark, and diboson background processes are found to be negligible from studies performed with simulated events.
The multijet background was studied with three different leading order Monte Carlo samples.
The \Pythia 8.186 event generator was used together with the A14 tune and the NNPDF2.3LO parton distribution functions,
 while the \Herwigpp\ 2.7.1 event generator was used together with the UEEE5 tune~\cite{Herwigpp} and CTEQ6L1 PDF sets~\cite{Pumplin:2002vw}.
The \Sherpa~event generator~\cite{Schumann:2007mg} was also used to generate multijet events for the study of background estimation.
Matrix elements were calculated with up to three partons 
at LO, were showered with \Sherpa~as well, and were merged using the ME+PS@LO prescription~\cite{Hoeche:2009rj}.
The CT10 PDF set~\cite{Lai:2010vv} was used in conjunction with dedicated parton shower tuning developed by the \Sherpa authors. 
For the generation of fully hadronic decays of \ttbar events, the \textsc{Powheg-Box v2}
event generator~\cite{Alioli:2010xd} was used with the CT10 PDF set and was interfaced with 
\Pythia 6.428~\cite{pythia}.
The EvtGen v1.2.0 program~\cite{Lange:2001uf} was also used to describe the properties of the $b$- and $c$- hadron
decays for the background samples except those generated with \Sherpa~\cite{Gleisberg:2008ta}.

The effect of additional $p$--$p$ interactions per bunch crossing (``pile-up'') as a function of the instantaneous
luminosity was taken into account by overlaying simulated minimum-bias events according to the observed
distribution of the number of pile-up interactions in data.
All Monte Carlo simulated background samples were passed
through a full $\textsc{Geant}4$ simulation~\cite{Agostinelli:2002hh} of the ATLAS
detector~\cite{SOFT-2010-01}. 
The signal samples were passed through a fast detector simulation~\cite{ATL-PHYS-PUB-2010-013} 
based on a parameterization of the performance of the ATLAS electromagnetic and hadronic calorimeters
and on $\textsc{Geant}4$ elsewhere. The compatibility of the signal selection efficiency between the fast simulation sample and the full simulation sample 
was validated at a number of signal points in the \sixQ\ and \tenQ\ considered in this Letter. 

\section{Event selection}
\label{sec:selection}

The data were recorded in 2015 and 2016, with the LHC operating at a centre-of-mass energy of $\sqrt{s} = 13\TeV$.
All detector elements are required to be operational.
The integrated luminosity is measured to be \lumiA~and \lumiB, for the 2015 and 2016 data sets, respectively. 
The uncertainty in the combined 2015 and 2016 integrated luminosity is 2.1\%. It is derived, following a methodology similar to that detailed in Ref.~\cite{DAPR-2013-01}, from a  calibration of the luminosity scale using $x$--$y$ beam-separation scans.

The events used in this search are selected using an $H_{\mathrm{T}}$~trigger, seeded from a first-level jet trigger with an $E_\mathrm{T}$~threshold of 100 GeV, 
which requires the scalar sum of jet transverse energies  at the high level trigger to be greater than 1.0 TeV. This requirement is found to be fully efficient for signal regions considered in this Letter.
Events are required to have a primary vertex with at least two
associated tracks with transverse momentum (\pT{}) above $0.4\GeV$.
The primary vertex assigned to the hard-scattering collision is the one
with the highest $\sum_\mathrm{track} \pT^2$, where the sum of
track $\pT^2$ is taken over all tracks associated with that vertex.
To reject events with detector noise or non-collision backgrounds, events are removed if they fail basic quality criteria~\cite{DAPR-2012-01,ATLAS-CONF-2015-029}.

Jets are reconstructed from three-dimensional topological clusters of energy deposits in the calorimeter
calibrated at the EM scale~\cite{PERF-2014-07}, using the \antikt~algorithm~\cite{Cacciari:2008gp,Cacciari:2011ma} with two different radius parameters of
$R$ = 1.0 and $R$ = 0.4, hereafter referred to as \largeR\ jets and \smallR\ jets, respectively. 
The four-momenta of the jets are calculated as the sum of the four-momenta of the
clusters, which are assumed to be massless. For the \largeR\ jets, the original constituents are calibrated
using the local cell weighting algorithm~\cite{ATL-LARG-PUB-2009-001,PERF-2014-07} prior to jet-finding and reclustered using the longitudinally-invariant \kt~algorithm~\cite{Catani:1993hr} with a radius
parameter of $R_{\text{sub-jet}}$ = 0.2, to form a collection of sub-jets.
 A sub-jet is discarded if it carries less than
5\% of the \largeR\ jet \pt~of the original jet. The constituents in the remaining sub-jets are then used to recalculate
the \largeR\ jet four-momenta, and the jet energy and mass are further calibrated to particle level using
correction factors derived from simulation~\cite{PERF-2012-02}. The resulting “trimmed”~\cite{Krohn:2009th,PERF-2012-02} \largeR\ jets are required to
have \pt > 200 GeV and |$\eta$| < 2.0. 
The analysis does not place any requirement on the
vertex association of tracks within a jet nor on the timing of the
calorimeter cells within a jet, which preserves the sensitivity of this
analysis to models containing non-prompt jets.
The \smallR\ jets are corrected for pile-up contributions and are then calibrated to the particle level using simulated events followed by a correction based on in situ measurements
~\cite{PERF-2016-04,ATL-PHYS-PUB-2015-036,PERF-2014-07}.

The identification of jets containing $b$-hadrons is based on the \smallR\ jets with \pT > 50 GeV and $|\eta|$ < 2.5
and a multivariate tagging algorithm~\cite{PERF-2012-04,ATL-PHYS-PUB-2016-012}. 
This algorithm is applied to a set of tracks with loose impact parameter constraints in a region of interest around each jet axis to enable
the reconstruction of the $b$-hadron decay vertex. The $b$-tagging requirements result in an efficiency of
70\% for jets containing $b$-hadrons, as determined in a sample of simulated \ttbar~events~\cite{ATL-PHYS-PUB-2016-012}.
A \smallR\ jet passing the $b$-tagging requirement is referred to as a $b$-tagged jet. 

The analysis of data is primarily based on observables built from \largeR\ jets. 
The \smallR\ jets are used to classify events and for categorization of the \largeR\ jets based on the $b$-tagging information. 
Specifically, events selected in the analysis are divided into a \btag sample where at least one $b$-tagged jet is present in the event, 
and a \bveto sample where no $b$-tagged jet is present in the event. Events selected without taking into account any $b$-tagging requirement are referred to as \binc events.
Large-$R$ jets are classified as either those that are matched to a $b$-tagged jet within $\DeltaR$ = 1.0 ($b$-matched jets), 
 or those that are not matched to a $b$-tagged jet.

\section{Analysis strategy}
\label{sec:strategy}

The analysis uses a kinematic observable, the total jet mass, \MJ~\cite{Hook:2012fd, Hedri:2013pvl, Cohen:2014epa},  
as the primary discriminating variable to separate signal and background.
The observable \MJ is defined as the sum of the masses of the four leading 
\largeR\ jets.

\begin{equation}
  \MJ = \sum_{{{\pT}>200\,\textrm{GeV} \atop |\eta|\leq 2.0} \atop j = 1 - 4} m_{\text{jet}}^{j} 
  \label{eq:defMJ}
\end{equation}

\begin{figure}[!htb]
  \centering
  \subfigure[]{
   \includegraphics[width=0.48\textwidth]{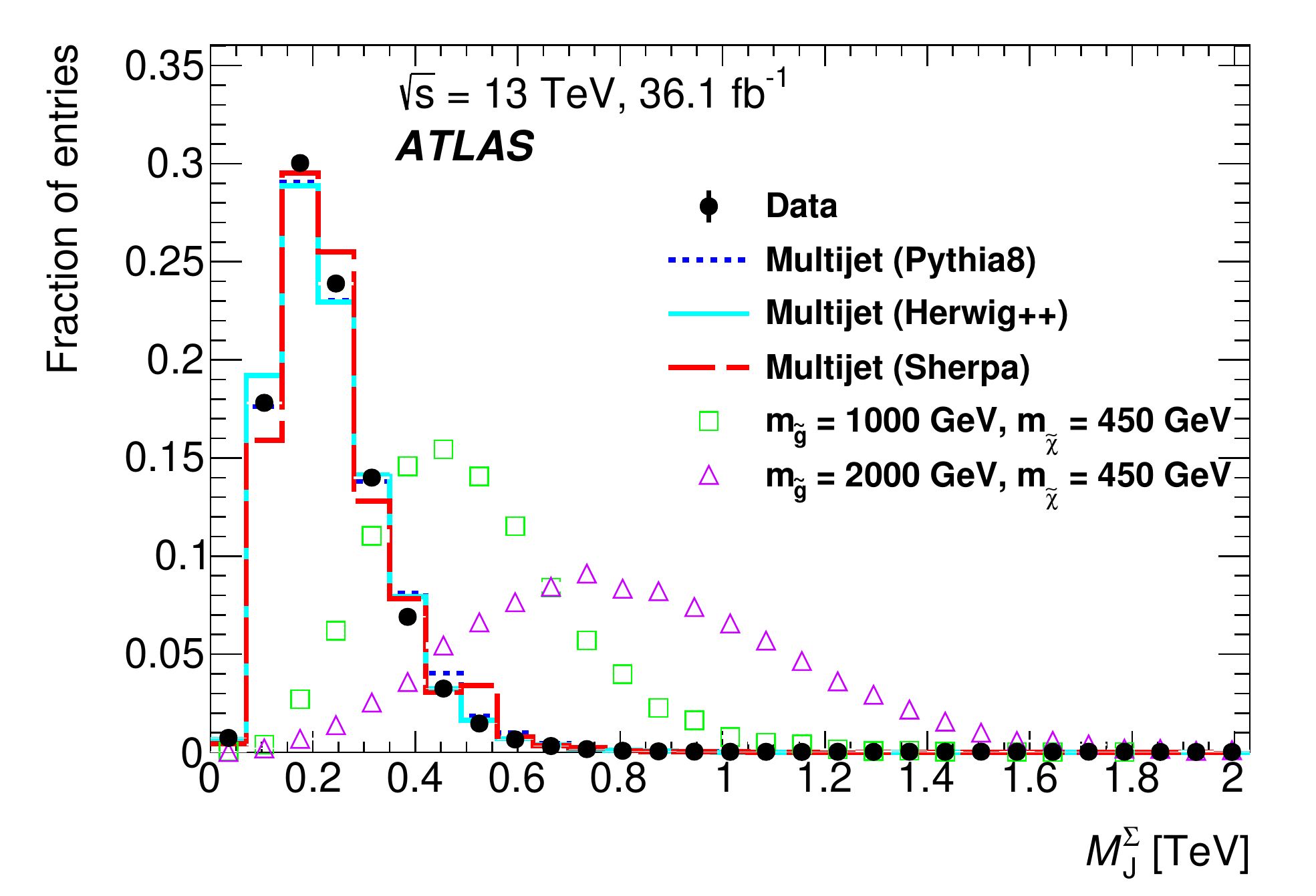}
    \label{fig:substructure:SvsB:MJ}}
  \subfigure[]{
   \includegraphics[width=0.48\textwidth]{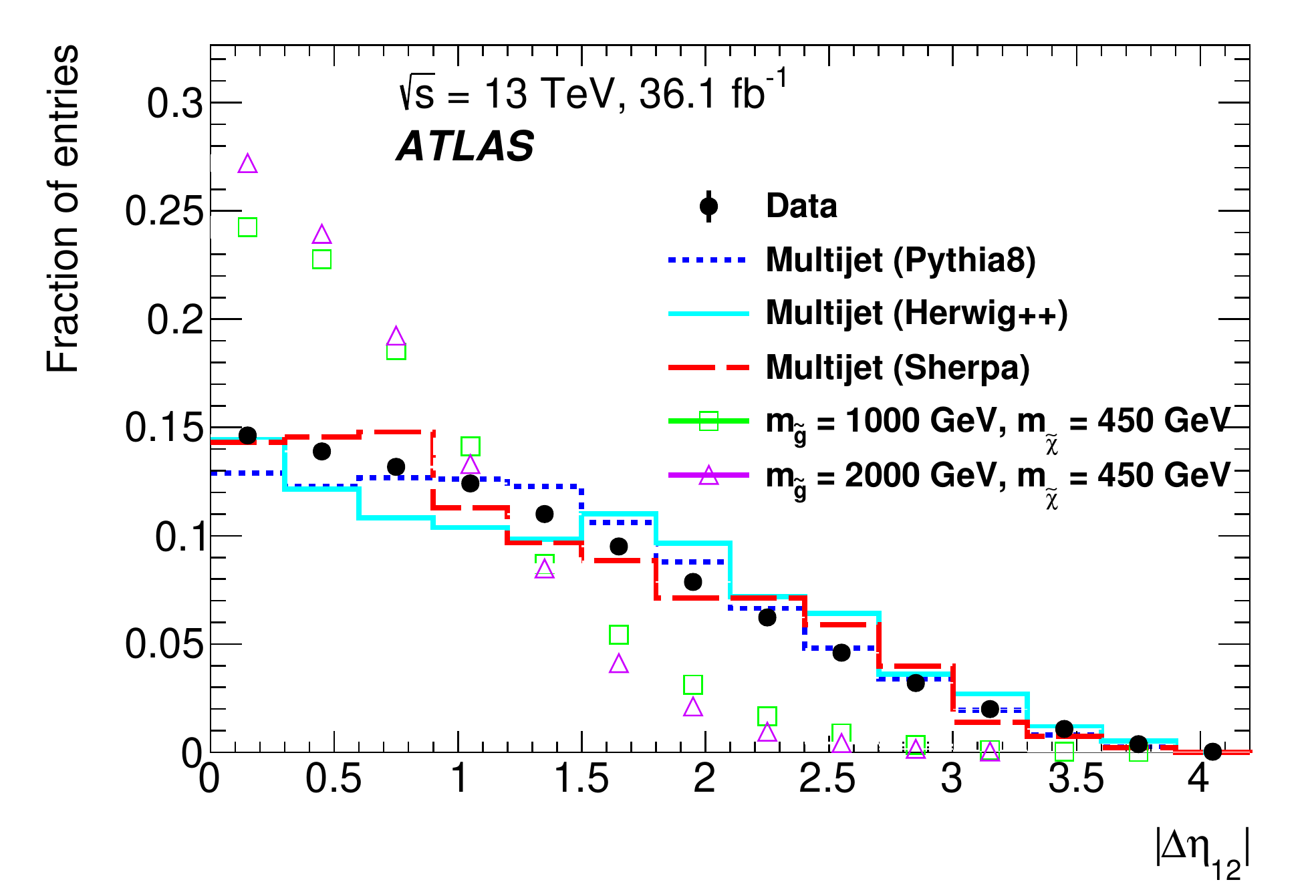}
    \label{fig:substructure:SvsB:DEta}}
  \caption{Comparison between signal samples and background control samples for \subref{fig:substructure:SvsB:MJ} the sum of the masses of the four leading \largeR jets \MJ and \subref{fig:substructure:SvsB:DEta} the difference in pseudorapidity between the two leading \largeR jets \deta. Two typical signal points for gluino cascade decay models are shown, as well as the distributions obtained from the data. All distributions are normalized to the same area. The selection requires four or more jets, is inclusive in \deta and has no $b$-tagging requirements.
           }
  \label{fig:substructure:SvsB}
\end{figure}

\noindent This observable provides significant sensitivity for gluinos with very high mass.
Figure~\ref{fig:substructure:SvsB:MJ}
presents examples of the discrimination that the \MJ observable provides between the background (represented here by \Sherpa, \Pythia\ 8.186 and \Herwigpp\ multijet Monte Carlo simulation) and several signal samples, as well as the comparison of the data to the simulated multijet background.

Another discriminating variable that is independent of \MJ is necessary in order to define suitable control and validation regions where the background estimation can be studied and tested.
The signal is characterized by a higher rate of central-jet events as compared to the primary multijet background.
This is expected due to the difference in the production modes: predominantly $s$-channel for the signal,
whereas the background can also be produced through $u$- and $t$-channel processes.
Figure~\ref{fig:substructure:SvsB:DEta} shows the distribution of the pseudorapidity difference between the two leading \largeR\ jets, \deta
 for several signal and background Monte Carlo samples, as well as data.
A high-\deta requirement can be applied to establish a control region or a validation region 
where the potential signal contamination needs to be suppressed.

The use of \MJ in this analysis provides an opportunity to employ the fully data-driven jet mass \textit{template method}
 to estimate the background contribution in signal regions. 
The jet mass template method is discussed in Ref.~\cite{Cohen:2014epa}, 
and its first experimental implementation is described in Ref.~\cite{SUSY-2013-07}.
In this method,
single-jet mass templates are extracted from signal-depleted control regions.
These jet mass templates are created in bins that are defined by a number of observables, which include jet \pt\ and $|\eta|$, and the $b$-matching status.
They provide a \textit{probability density function} that describes the relative probability for a jet with a given \pT\ and $\eta$ to have a certain mass.
This method assumes that jet mass templates only depend on these observables 
 and are the same in the control regions and signal regions.
A sample where the background \MJ~distribution needs to be estimated, such as a validation region or a signal region, 
 is referred to as the kinematic sample.
The only information used is the jet \pT\ and $\eta$, as well as its $b$-matching status,
which are inputs to the templates.
For each jet in the kinematic sample, 
 its corresponding jet mass template is used to generate a random jet mass. 
An \MJ distribution can be constructed from the randomized jet masses of the kinematic sample. 
If jet mass templates are created from a control sample of background events, then the \MJ~distribution constructed from randomized jet masses should reproduce the shape of the \MJ distribution for the background.\footnote{When signal events are present in the kinematic sample, a correction is needed in order to remove the bias in the background estimate, and this correction is discussed later in this letter.}

This jet mass prediction procedure is similar to the one employed in Ref.~\cite{SUSY-2013-07} with two minor differences.
First, the statistical fluctuations in the jet mass templates are propagated to the background yield prediction in the signal region, 
and therefore considered as a systematic uncertainty of the jet mass template method, 
whereas the Run-1 analysis made assumptions about the form of the template shape by smoothing using a Gaussian kernel technique.
Second, the predicted \MJ distribution is normalized to the observation in 0.2~TeV < \MJ < 0.6 TeV, 
whereas the Run-1 analysis did not introduce any normalization region, effectively normalizing the prediction to the observation in the entire \MJ range.
The boundaries of the normalization region are determined so that contamination from signal models not yet excluded by the previous search~\cite{SUSY-2013-07} 
is negligible compared to the statistical uncertainty of the background.
  
\begin{figure}[!htb]
  \centering
  \subfigure[]{
   \includegraphics[width=0.48\textwidth]{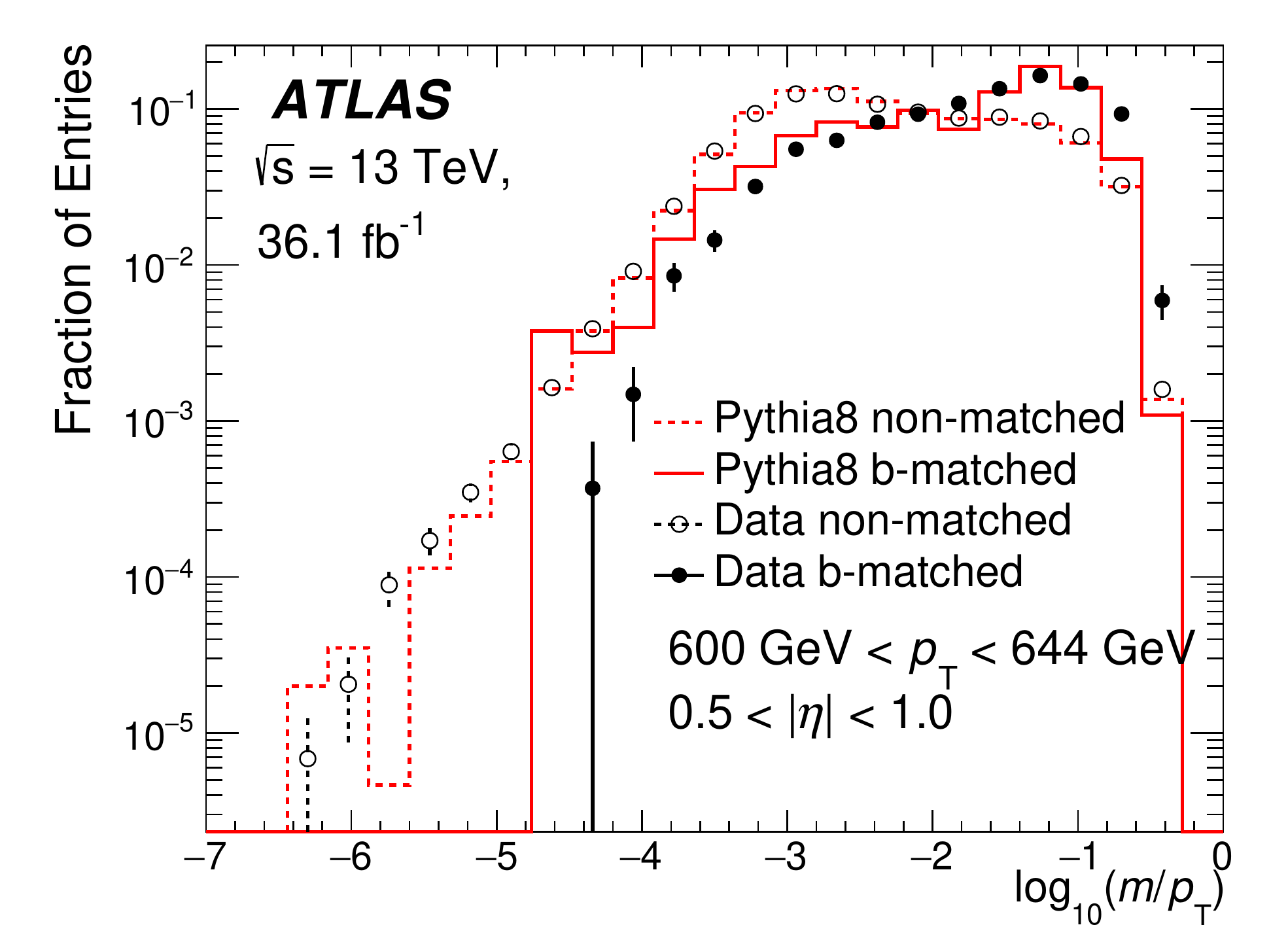}
    \label{fig:template:bin16}}
  \subfigure[]{
   \includegraphics[width=0.48\textwidth]{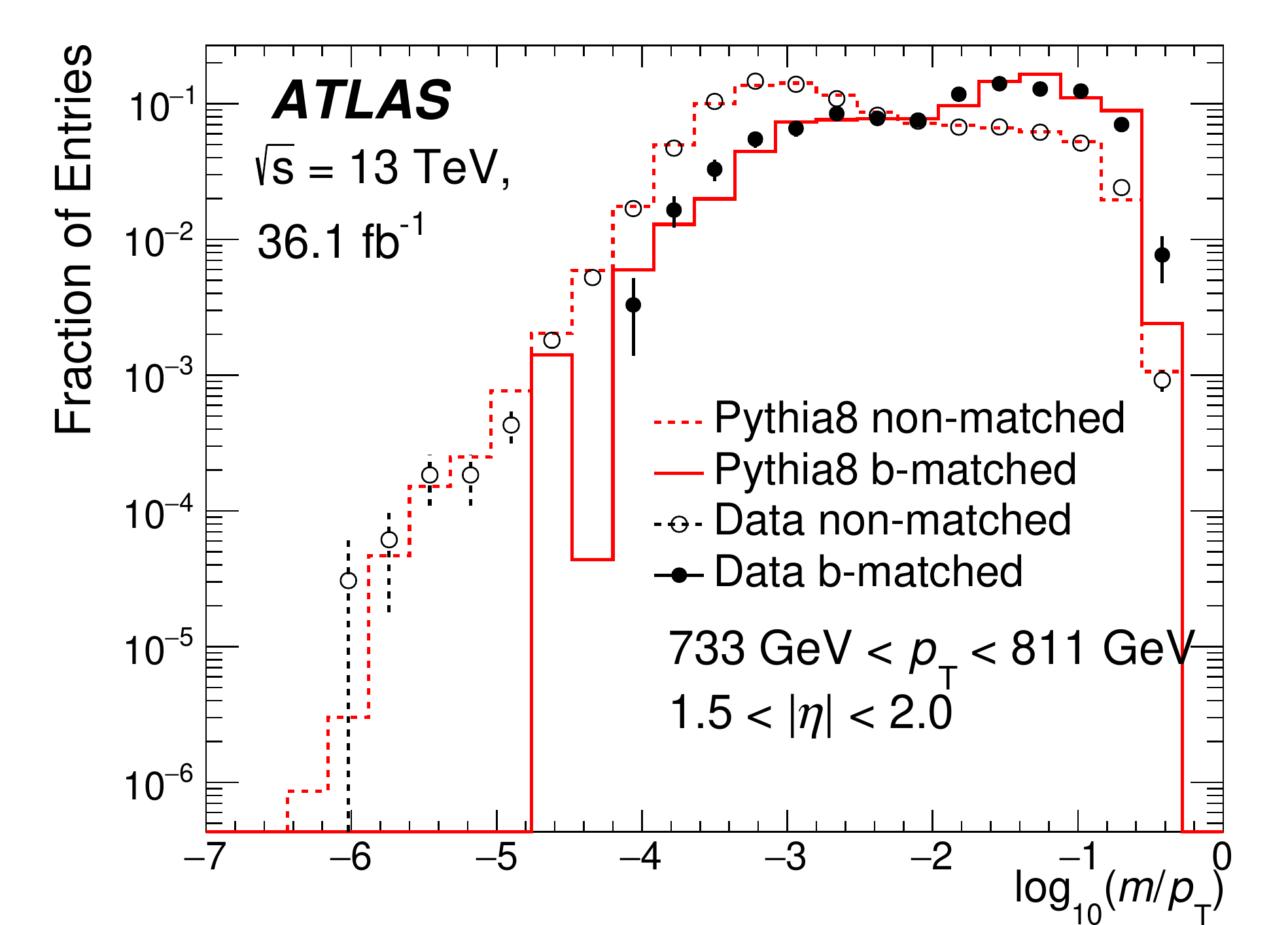}
    \label{fig:template:bin23}}
  \caption{ Example jet mass template distributions for $b$-matched jets and non-matched jets in data (solid and open circles) and 
\PythiaE\ multijet (solid and dashed lines) samples.
 ~\subref{fig:template:bin16} shows the jet mass template distributions in the bin of 
600~GeV < \pt\ < 644 GeV, 0.5 < $|\eta|$ < 1.0, while~\subref{fig:template:bin23} shows the jet mass template distributions 
in the  bin of 733~GeV < \pt\ < 811 GeV, 1.5 < $|\eta|$ < 2.0.           }
  \label{fig:template}
\end{figure}

The selected events are divided into control, uncertainty determination, validation and signal regions, as summarized in Table~\ref{tab:regions}.
Control regions (CRs) are defined with events that have exactly three \largeR\ jets with \pT > 200 GeV.
Jets in the control regions are divided into 4 $|\eta|$ bins uniformly defined between 0 and 2, 15 $\pt$ bins uniformly defined in $\mathrm{log_{10}}(\pt)$,
and 2 $b$-matching status bins ($b$-matched or not). A total of 120 jet mass templates are created.
Figure~\ref{fig:template} shows example jet mass template distributions in two \pt--$|\eta|$ bins 
 for both the data and \PythiaE\ multijet samples. The shapes of the jet mass templates are different 
between $b$-matched jets and non-matched jets.
A \deta~> 1.4 requirement is included for control region events where at least one $b$-matched jet is present, in order to suppress potential signal contamination.

Five overlapping signal regions (SRs) are considered in this analysis.
All signal regions are required to have \deta < 1.4.
The first set of signal regions does not require the presence of a $b$-tagged jet 
and is used to test more generic BSM signals of pair-produced heavy particles cascade-decaying into many quarks or gluons.
Two selections on the \largeR\ jet multiplicity are used, \njet $\ge$ 4 (4jSR) and \njet $\ge$ 5 (5jSR).
In order to further improve the sensitivity to the benchmark signal models of the RPV UDD scenario,
subsets of events in the 4jSR and 5jSR are selected by requiring the presence of at least one \btagged \smallR\ jet.
To ensure that the $H_{\mathrm{T}}$ trigger is fully efficient for the offline data analysis, 
a leading-jet \pt~> 400 GeV requirement is added for signal regions with four or more \largeR\ jets.
Finally, a requirement on the \MJ\ variable is placed in each signal region,
 with the requirement optimized for the direct decay and cascade decay models. 
For each signal region, a validation region is defined by reversing the \deta\ requirement. These validation regions are used to cross-check the background estimation,
thus validating the background prediction in the signal region.

\begin{table}[t]
\caption{Summary of the event-level and jet-level requirements used to define various regions.
Requirements on \largeR\ jet multiplicity (\njet), whether or not a $b$-tagged jet is present ($b$-tag), and the pseudorapidity gap between the two leading-\largeR-jets (\deta) are applied to
 define control, validation and signal regions. In addition, each signal region includes an additional \MJ\ requirement for statistical interpretation. 
Control regions are defined separately for non-matched jets and $b$-matched jets.
For the uncertainty determination regions, the \njet~and leading-jet \pt\ ($p_{\mathrm{T},1}$) requirements are used.
}
  \begin{center}
    \begin{tabular}{c c c c c c c}
\hline
 \multicolumn{2}{c}{}          & \njet~(\pt\ > 200 GeV) & $b$-tag  & $p_{\mathrm{T},1}$ & \deta & \MJ\\
\hline
CR                     & 3jCR    &  = 3   &  -   & -         &  - & - \\
\hline
 \multirow{2}{*}{UDR} & UDR1   & = 2   &  -        &  > 400 GeV         & - & -\\
                     & UDR2    &  = 4   &  -         & < 400 GeV         &  - & - \\
\hline
 \multirow{4}{*}{VR} & 4jVR   & $\ge$ 4   &  -       & > 400 GeV         &  > 1.4& -\\
                     & 5jVR   & $\ge$ 5   &  -         & -         &  > 1.4& -\\
                     & 4jVRb   & $\ge$ 4   &  $\ge$~1      & > 400 GeV         & > 1.4 & -\\
                     & 5jVRb   & $\ge$ 5   &  $\ge$~1        & -         &  > 1.4& -\\
\hline
 \multirow{4}{*}{SR} & 4jSR   & $\ge$ 4   &  -         & > 400 GeV         &  < 1.4 & > 1.0 TeV\\
                     & 5jSR   & $\ge$ 5   &  -         & -         &  < 1.4 & > 0.8 TeV\\
                     & 4jSRb   & $\ge$ 4   &  $\ge$~1         & > 400 GeV         & < 1.4 & > 1.0 TeV\\
                     & 5jSRb\_1   & $\ge$ 5   &  $\ge$~1        & -         &  < 1.4 & > 0.8 TeV\\
                     & 5jSRb\_2                       & $\ge$ 5   &  $\ge$~1        & -         &  < 1.4 & > 0.6 TeV\\
\hline

\hline
\end{tabular}
\end{center}
\label{tab:regions}
\end{table}

Uncertainties in the jet mass prediction include a statistical component and a systematic component. 
The statistical uncertainty arises from the finite sample size in the control region, and the jet mass randomization,
 which can be quantified through pseudo-experiments.
Systematic uncertainties of the jet mass prediction can be attributed to a number of factors; for example,
jet mass templates are assumed to only depend on a given number of observables (jet \pt, $|\eta|$, and $b$-matching information, in this analysis),
jet mass templates are created for each of these observables with a given bin width, 
 and jets in the same event are assumed to be uncorrelated with each other, such that their masses can be modelled independently.
These systematic uncertainties are estimated in uncertainty determination regions (UDRs) in data, 
 where the predicted and observed jet masses are compared. 
The difference between them provides an estimate of the size of the systematic uncertainty. 

The UDRs represent extreme scenarios in terms of jet origin and multiplicity of an event, and the uncertainties estimated from these regions are found to be large enough to 
cover the potential difference between the true and estimated background in the signal regions. 
This strategy has been validated with the simulated background samples.
 One UDR (UDR1) requires exactly two \largeR\ jets with the leading \largeR\ jet \pt\ greater than 400 GeV. 
Events in this UDR contain high-\pt\ jets and can have an imbalance in \pt\ between the leading-jet and the subleading-jet. 
The other UDR (UDR2) is defined by requiring exactly four \largeR\ jets with the leading \largeR\ jet \pt\ less than 400 GeV.
Events in this UDR contain fewer energetic jets, which tend to be more balanced in \pt. 
In each UDR, selected jets are binned in the same way as they are in the control regions. 

In order to quantify the small difference between the predicted and observed jet mass distributions, 
 the jet mass response, defined as the ratio of the average observed jet mass to the average predicted jet mass, 
is studied with both UDRs.
It is found that the difference between jet mass distributions in the same \pt\ and $|\eta|$ bin between regions with different selections
can be largely captured by a scale factor between the distributions, and therefore the jet mass response
reflects the size of this scale factor.  Studies using Monte Carlo multijet events have shown that
scaling up and down the predicted jet mass by the jet mass response in the UDRs 
leads to variations in the predicted \MJ\ distributions that cover the difference between the observed and predicted \MJ\ distributions.

\begin{figure}[!htb]
  \centering
  \subfigure[]{
   \includegraphics[width=0.48\textwidth]{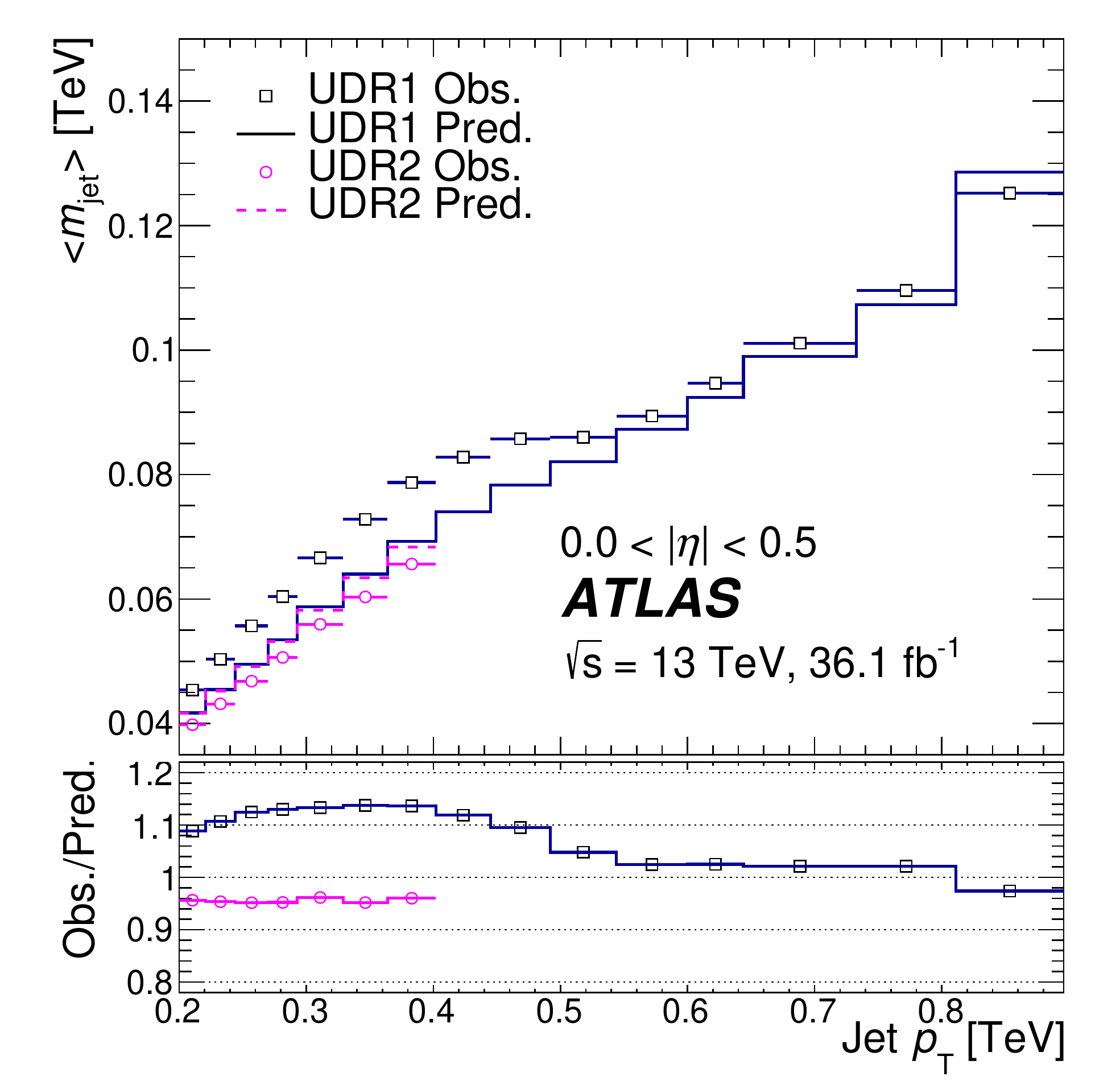}
    \label{fig:UDR:e1}}
  \subfigure[]{
   \includegraphics[width=0.48\textwidth]{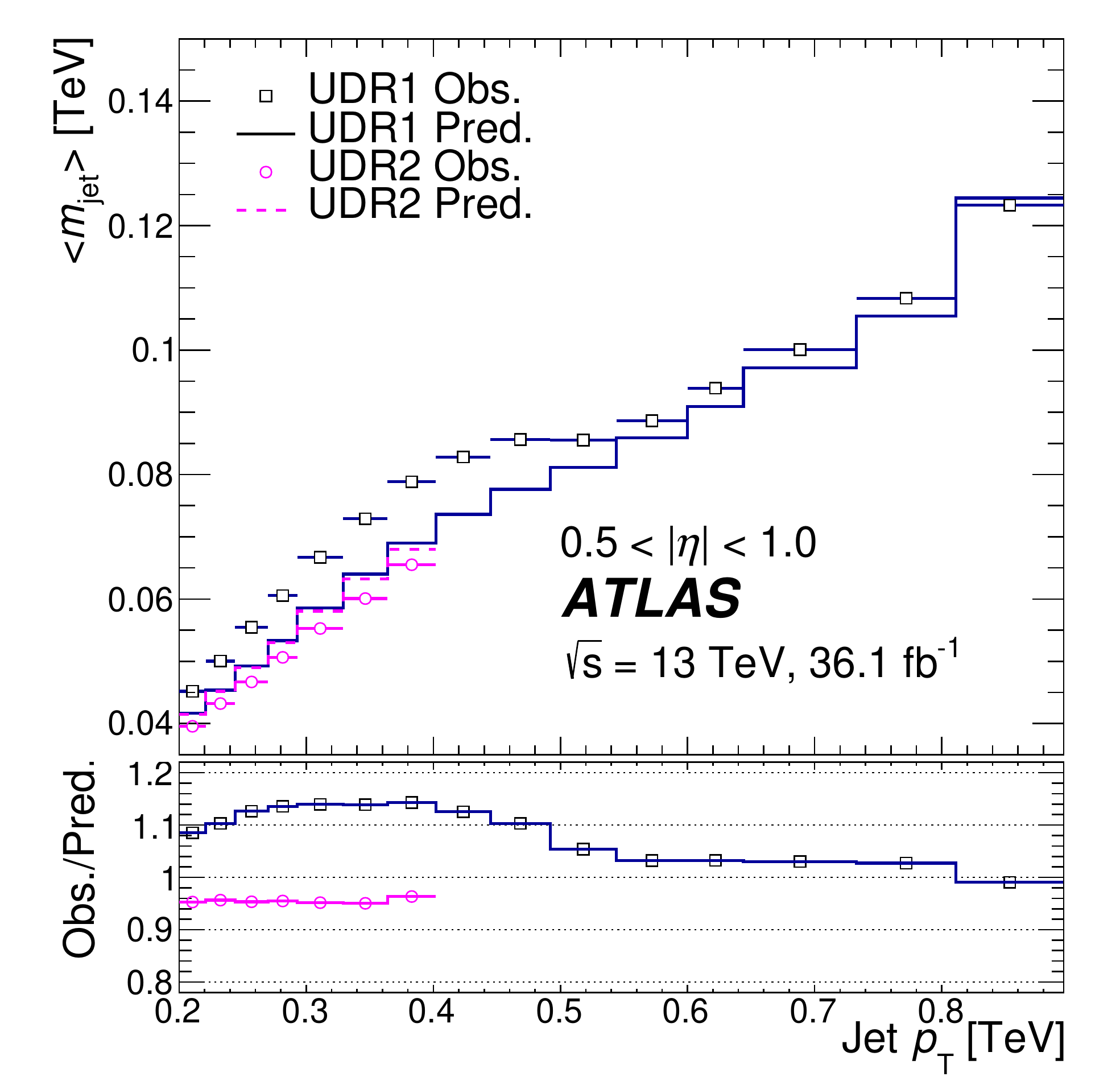}
    \label{fig:UDR:e2}} \\
  \subfigure[]{
   \includegraphics[width=0.48\textwidth]{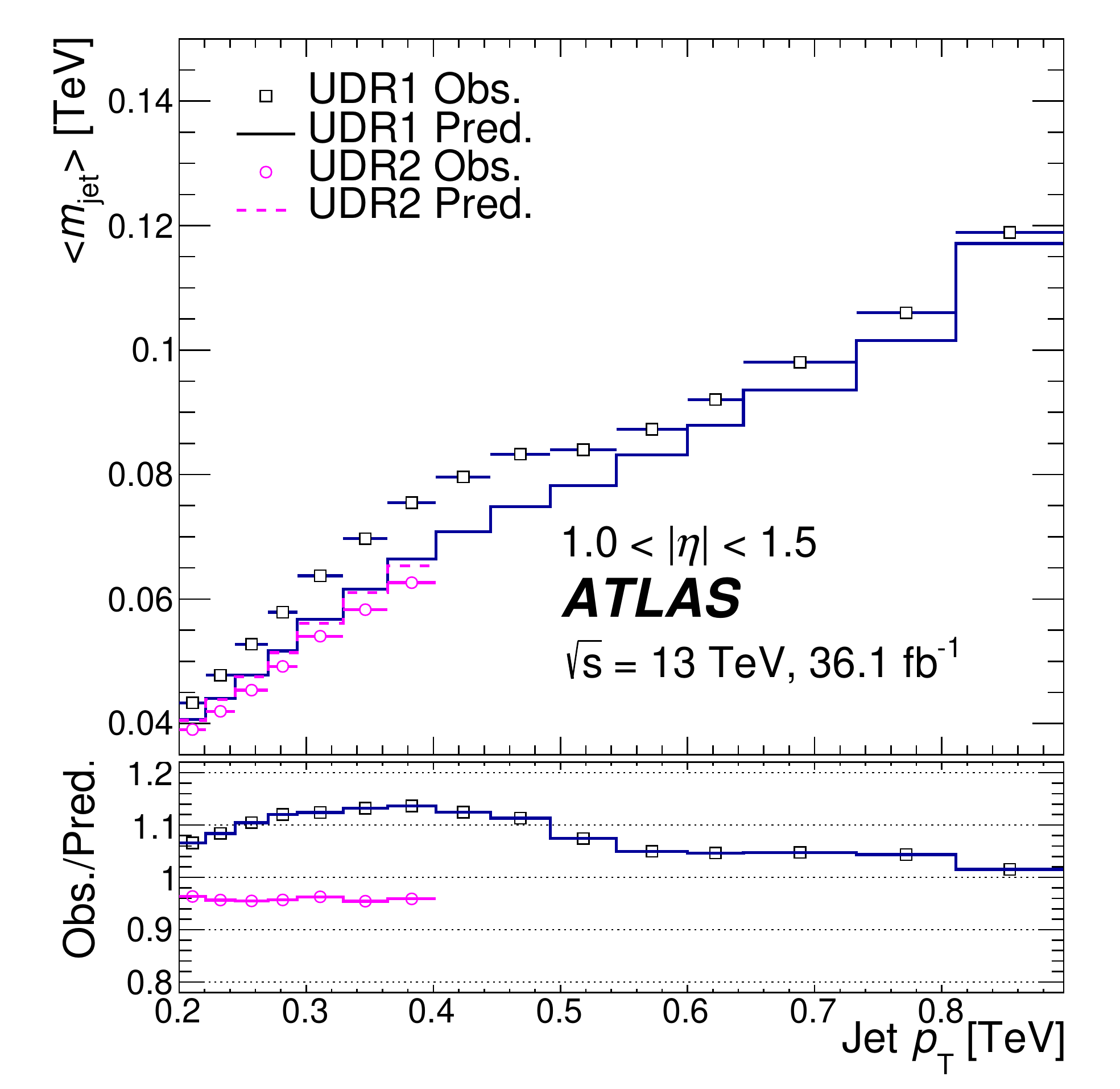}
    \label{fig:UDR:e3}}
  \subfigure[]{
   \includegraphics[width=0.48\textwidth]{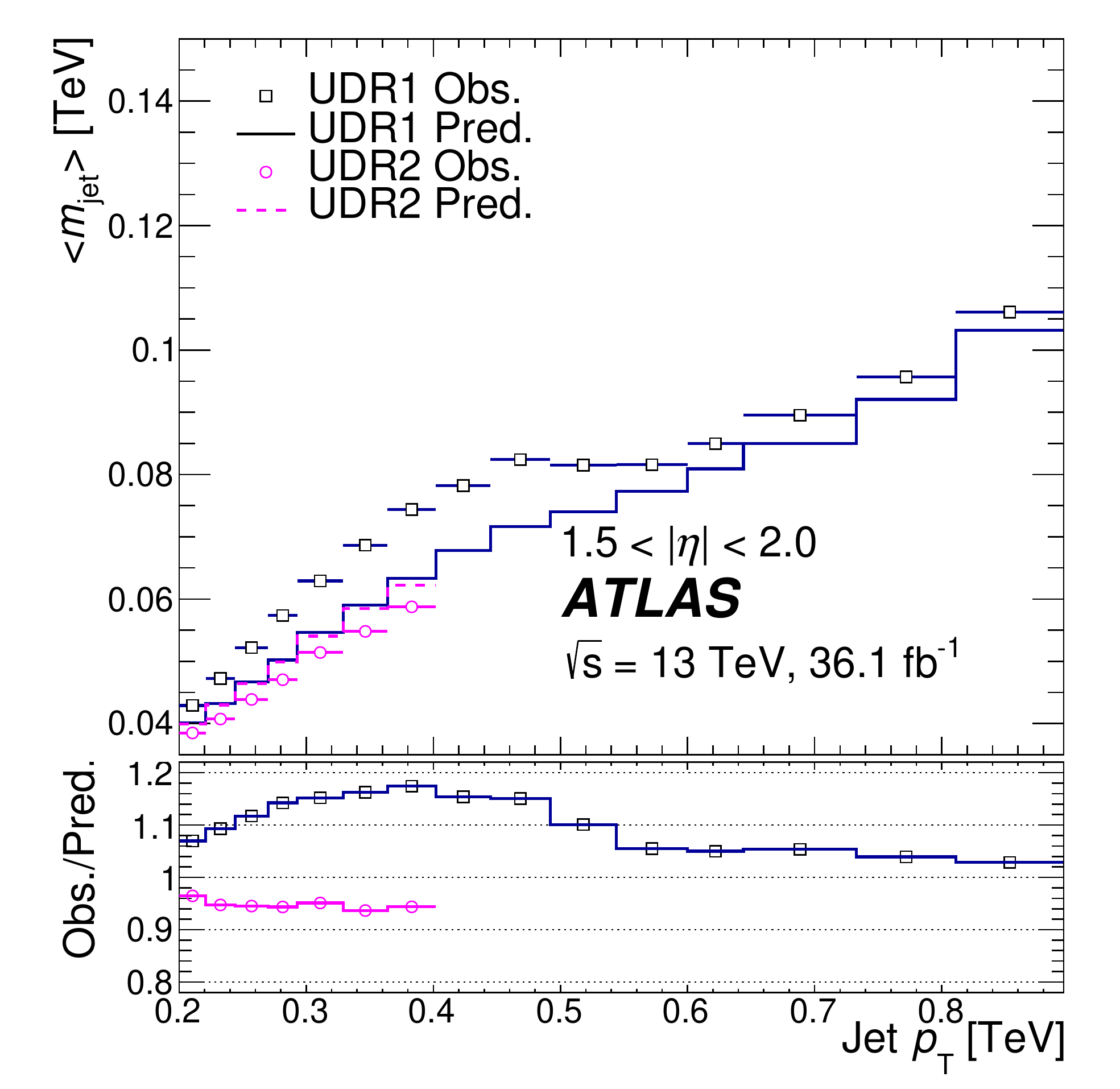}
    \label{fig:UDR:e4}} \\
  \caption{ The average observed and predicted jet masses (top panes)
 and the jet mass responses (bottom pane) in UDR1 and UDR2
 are shown for four different pseudorapidity regions.
           }
  \label{fig:UDR:response}
\end{figure}

Figure~\ref{fig:UDR:response} shows the jet mass responses in the UDRs as a function of jet \pt~and $|\eta|$.
An under-prediction of jet mass is seen in the UDR1, varying between a few percent and 14\%.
In the \pt\ range of 200 GeV--400 GeV, the UDR2 indicates an over-prediction, at the 4--5\% level.
Overall, the behaviour of the jet mass response is quite similar between different pseudorapidity regions.
It was checked and found that the difference between predicted and observed jet masses in the UDRs are not due to the trigger
inefficiency in the UDRs and CR, based on studies performed with Monte Carlo multijet samples and data. In these studies, additional $H_{\mathrm{T}}$~requirements are introduced in the analysis so that the UDRs and CR are fully efficient with respect to the HLT\_ht1000 trigger, and the differences in the UDRs remain qualitatively the same.
The differences in the jet mass response are used as an estimate for the \pt- and $|\eta|$-dependent systematic 
 uncertainty of the jet mass prediction.
Since the signs of the differences from the UDR1 and UDR2 are opposite in the \pt\ range of 200 GeV--400 GeV, 
the larger of the differences from these UDRs is used as the uncertainty and symmetrized. 
The uncertainty of the jet mass prediction is uncorrelated between the \pt\ range of 200 GeV--400 GeV (``low-\pt'')
 and the \pt\ range of > 400 GeV (``high-\pt''). For jets within the low-\pt\ or high-\pt\ range, the 
jet mass prediction uncertainties are correlated between different \pt\ and $|\eta|$ bins. 

\newpage

Possible bias on the background estimate due to the presence of \ttbar~events, where the jet origin is different from that in multijet events, 
is not explicitly addressed by the background estimation strategy.
However, a study using Monte Carlo multijet and \ttbar samples finds that the background prediction 
 is insensitive to the presence of \ttbar\ events, because of its relatively small cross section.
 
\begin{figure}[!htb]
  \centering
  \subfigure[$\ge$ 4 jets, \binc~events (4jVR)]{
   \includegraphics[width=0.48\textwidth]{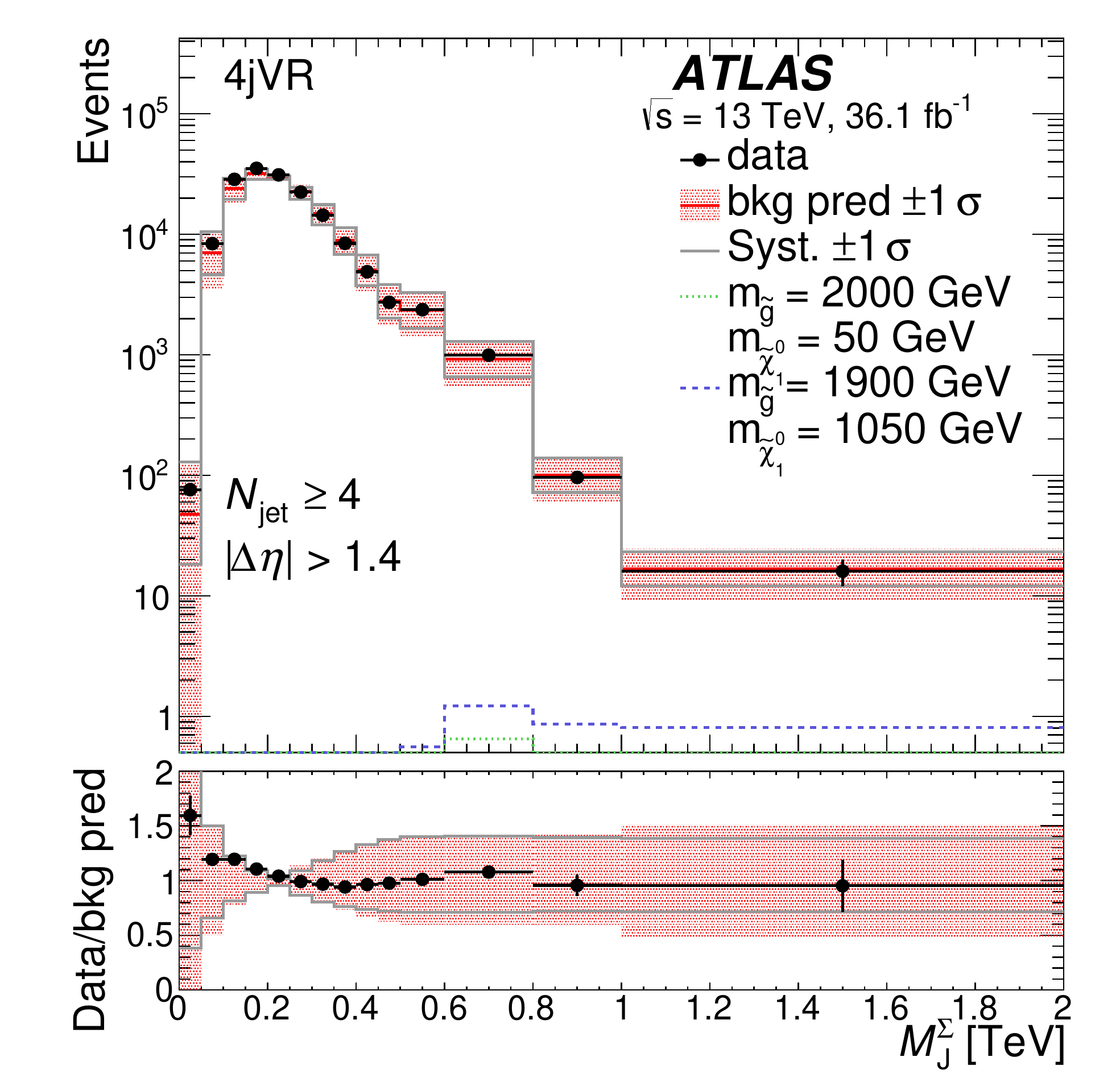}
    \label{fig:validation:4jVR}}
  \subfigure[$\ge$ 4 jets, \btag~events (4jVRb)]{
   \includegraphics[width=0.48\textwidth]{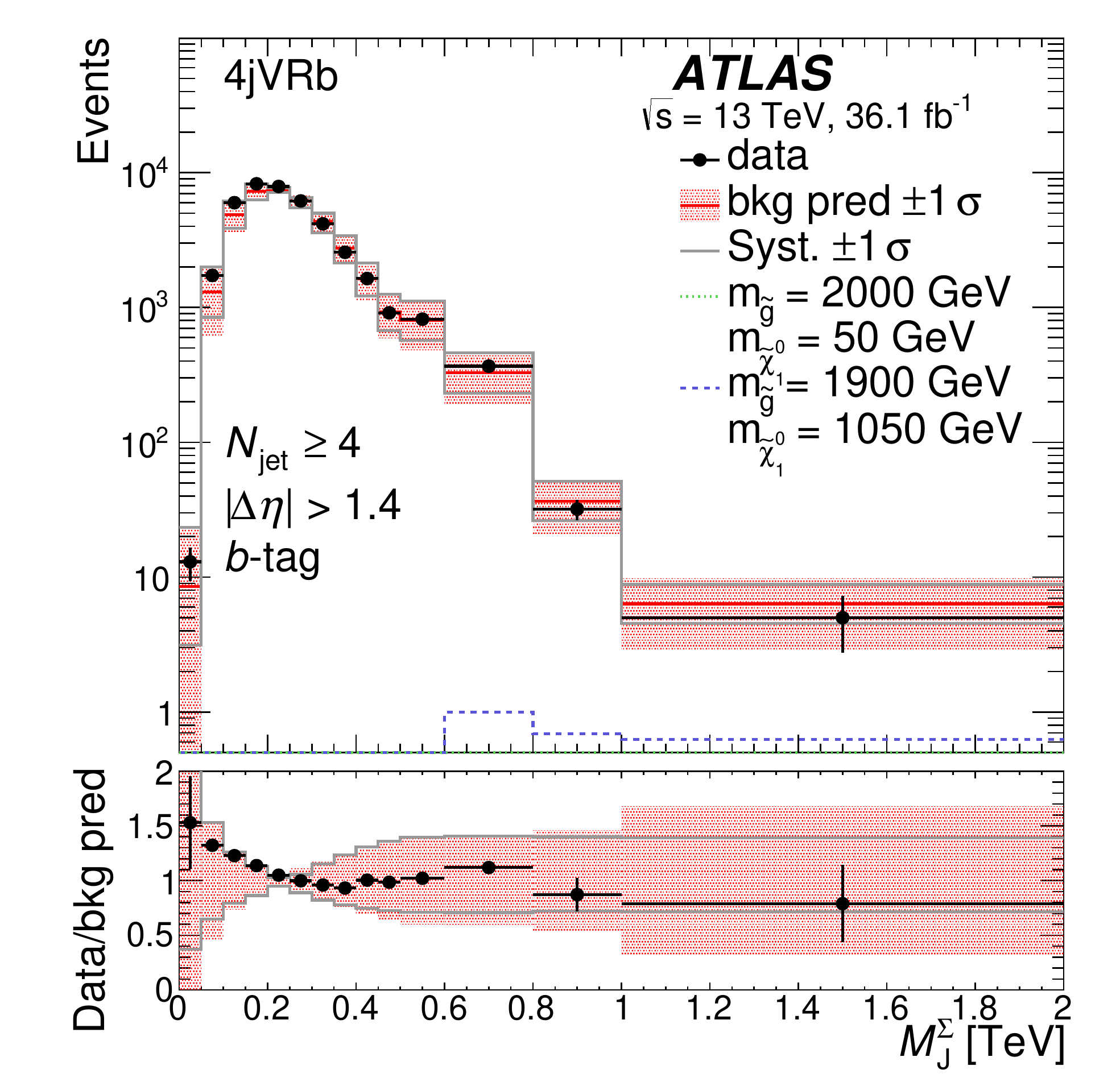}
    \label{fig:validation:4jVRb}} \\

  \subfigure[$\ge$ 5 jets, \binc~events (5jVR)]{
   \includegraphics[width=0.48\textwidth]{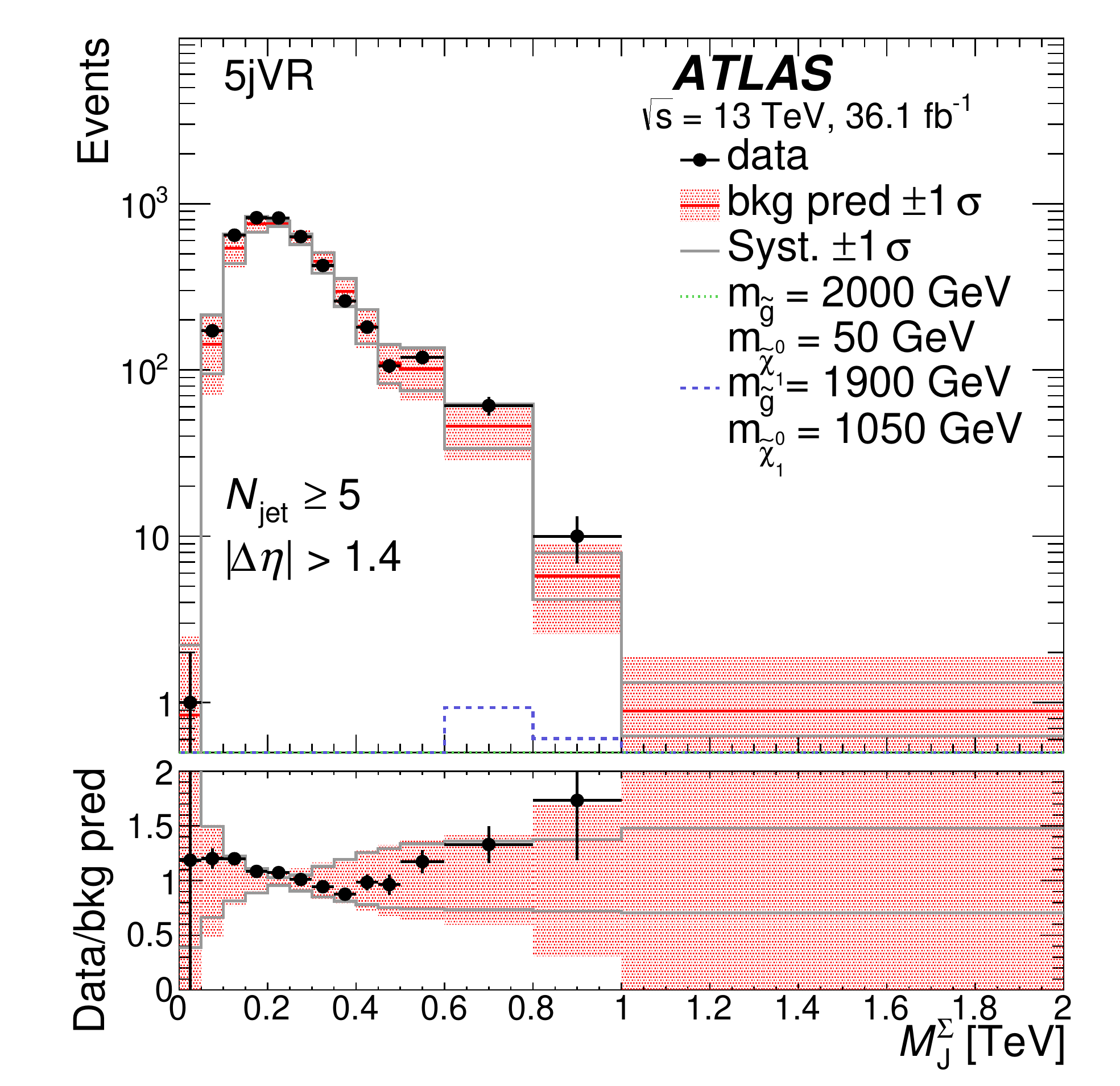}
    \label{fig:validation:5jVR}}
  \subfigure[$\ge$ 5 jets, \btag~events (5jVRb)]{
   \includegraphics[width=0.48\textwidth]{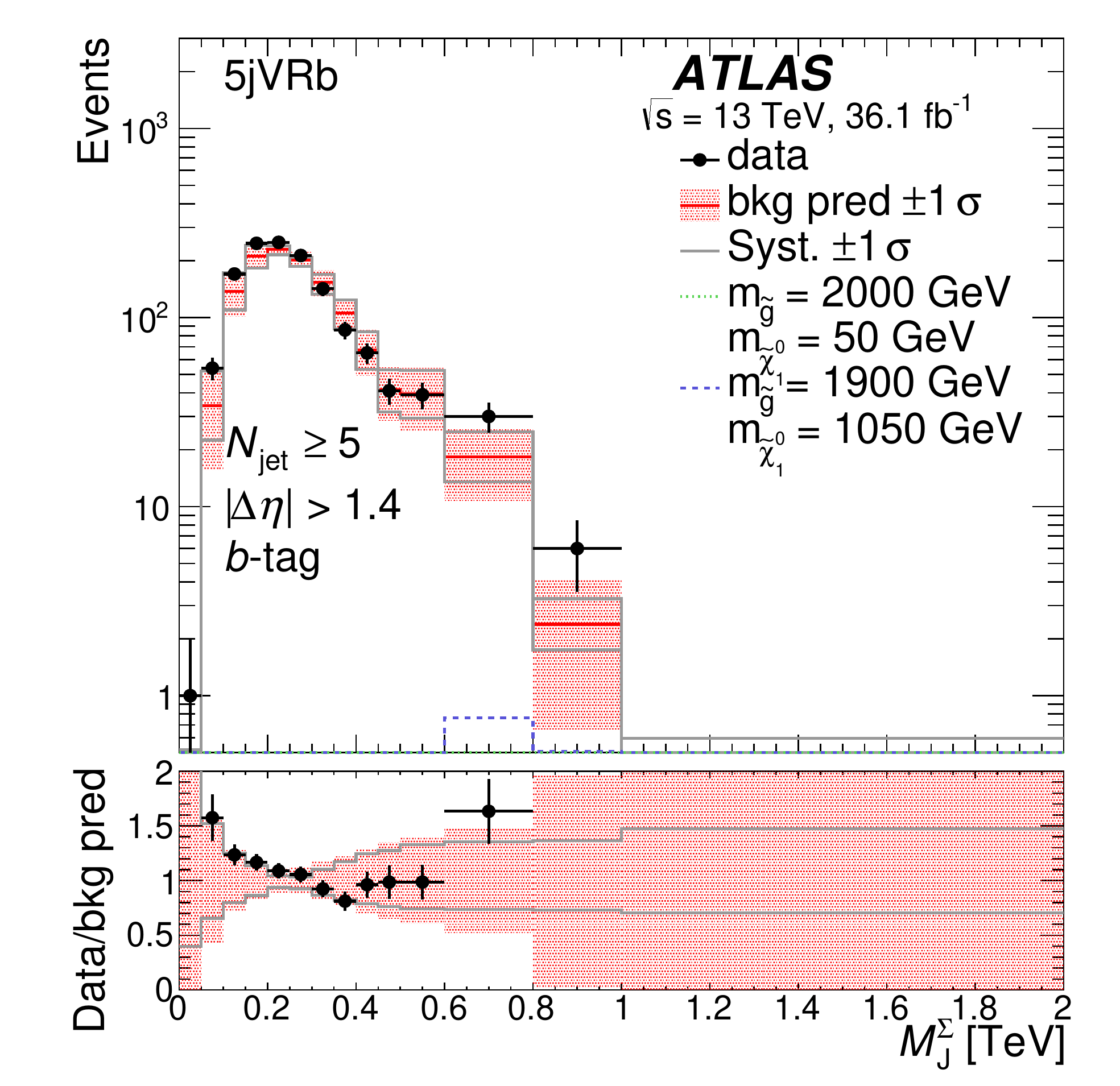}
    \label{fig:validation:5jVRb}} \\
  \caption{ Predicted (solid line) and observed (dots) \MJ distributions for validation regions \subref{fig:validation:4jVR} 4jVR, \subref{fig:validation:4jVRb} 4jVRb, 
\subref{fig:validation:5jVR} 5jVR, and \subref{fig:validation:5jVRb} 5jVRb. 
The shaded area surrounding the predicted \MJ distribution represents the uncertainty of the background estimation. 
The predicted \MJ distribution is normalized to data in 0.2 TeV < \MJ < 0.6 TeV,
where the expected contaminations from signals of \tenorsixQ not excluded by the Run-1 analysis~\cite{SUSY-2013-07}
are negligible compared to the background statistical uncertainty.
The expected contributions from two RPV signal samples are also shown.
           }
  \label{fig:validation:MJ}
\end{figure}

\begin{figure}[!htb]
  \centering

  \subfigure[$\ge$ 4 jets, \binc~events (4jSR)]{
   \includegraphics[width=0.48\textwidth]{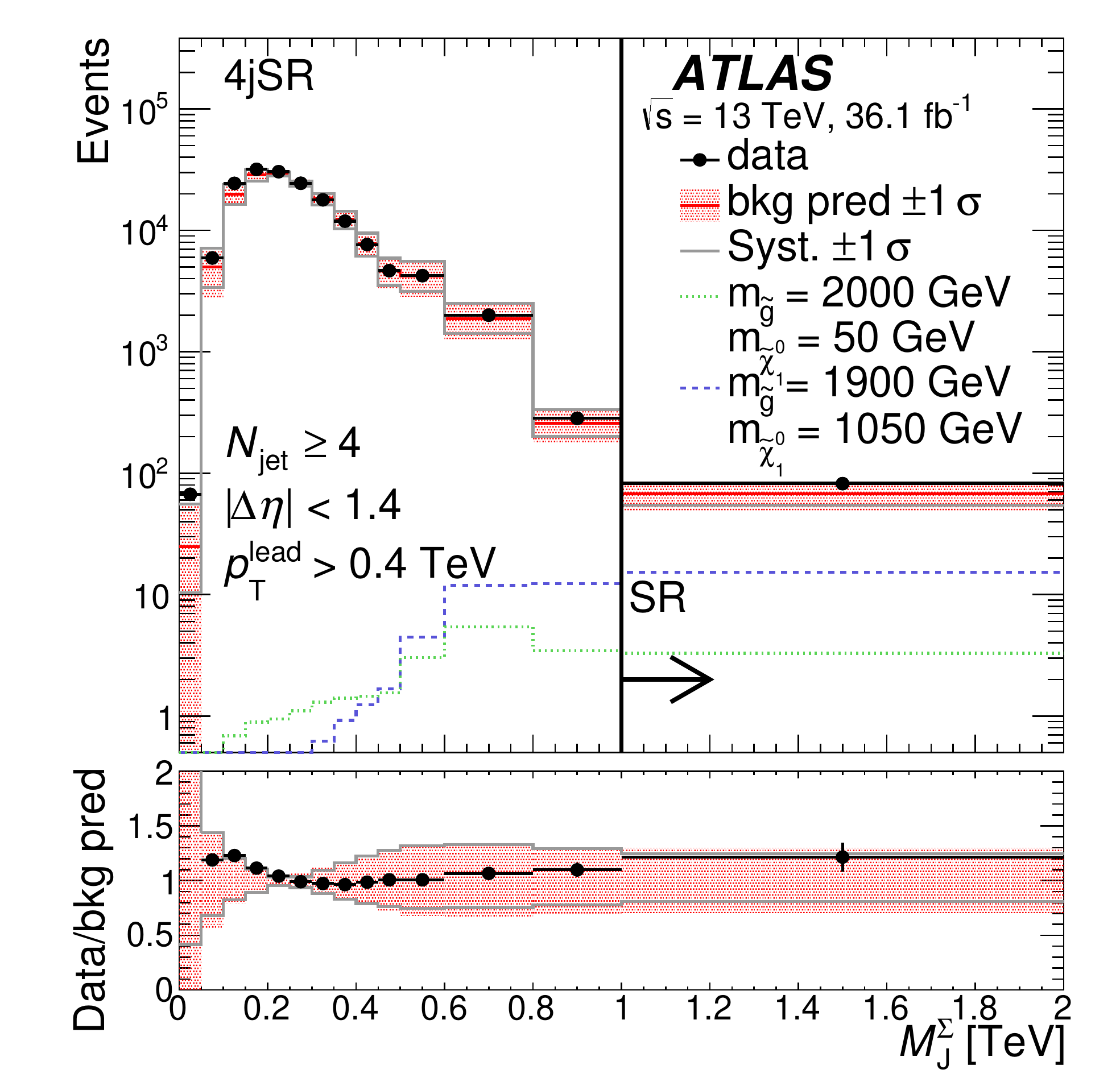}
    \label{fig:validation:4jSR}}
  \subfigure[$\ge$ 4 jets, \btag~events (4jSRb)]{
   \includegraphics[width=0.48\textwidth]{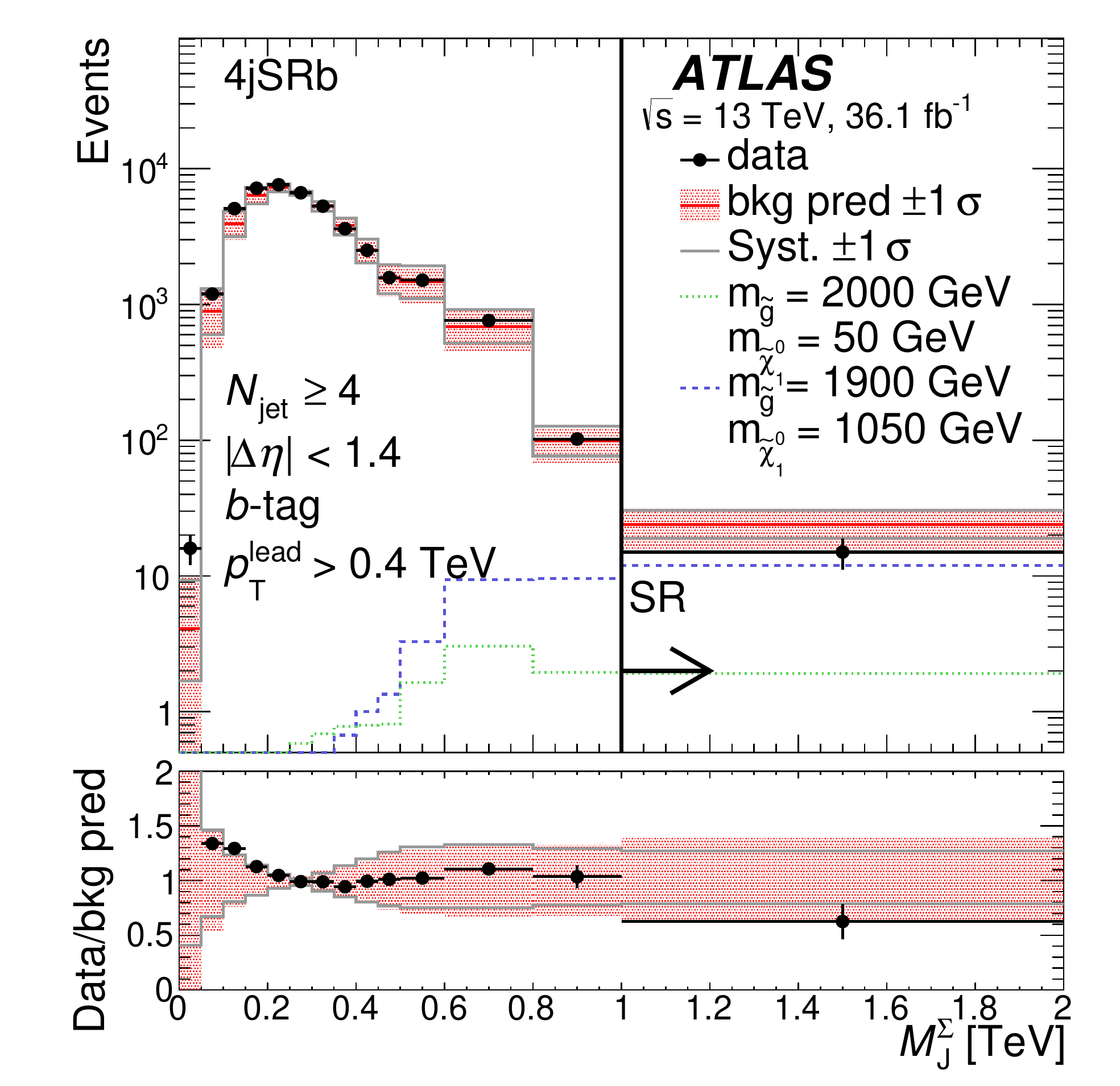}
    \label{fig:validation:4jSRb}} \\

  \subfigure[$\ge$ 5 jets, \binc~events (5jSR)]{
   \includegraphics[width=0.48\textwidth]{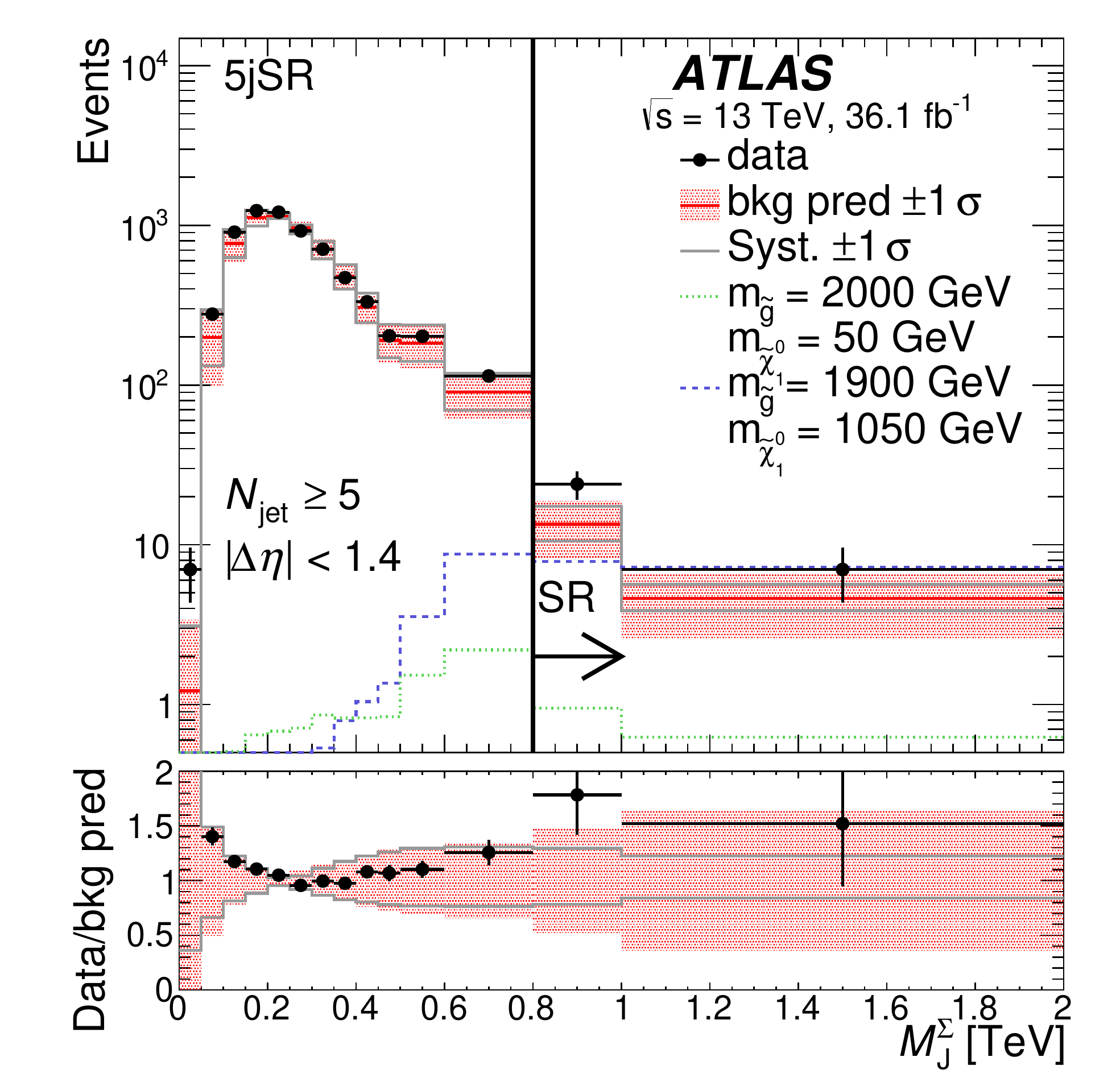}
    \label{fig:validation:5jSR}}
  \subfigure[$\ge$ 5 jets, \btag~events (5jSRb)]{
   \includegraphics[width=0.48\textwidth]{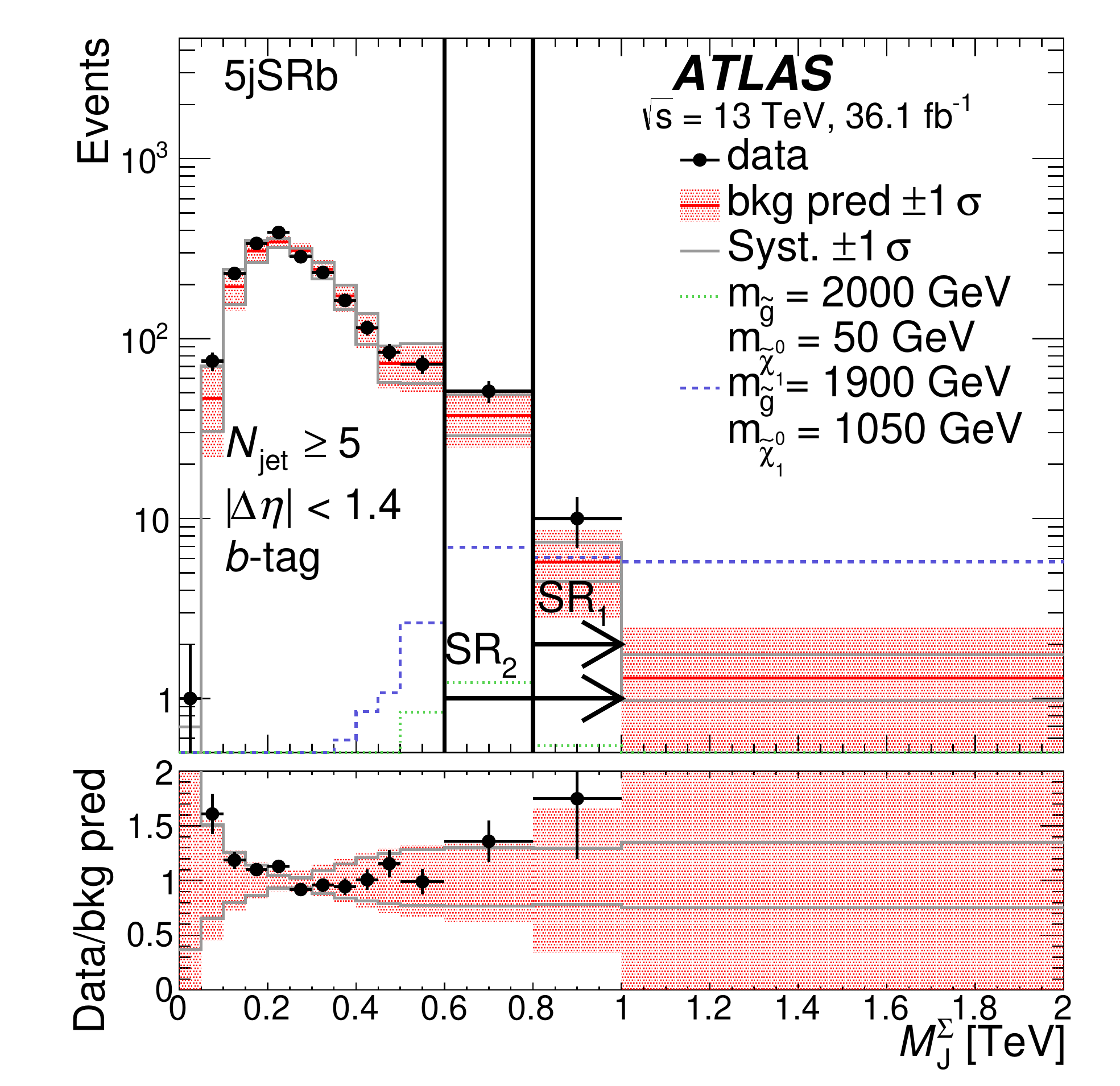}
    \label{fig:validation:5jSRb}} \\
  \caption{ Predicted (solid line) and observed (dots) \MJ distributions for signal regions \subref{fig:validation:4jSR} 4jSR, \subref{fig:validation:4jSRb} 4jSRb,
\subref{fig:validation:5jSR} 5jSR, and \subref{fig:validation:5jSRb} 5jSRb.
The shaded area surrounding the predicted \MJ distribution represents the uncertainty of background estimation.
The predicted \MJ distribution is normalized to data in 0.2 TeV < \MJ < 0.6 TeV, 
where the expected contaminations from signals of \tenorsixQ not excluded by the Run-1 analysis~\cite{SUSY-2013-07}
are negligible compared to the background statistical uncertainty. 
The expected contributions from two RPV signal samples are also shown.
           }
  \label{fig:signal:MJ}
\end{figure}

The jet mass template method is then applied to data in the validation and signal regions.
Uncertainties in the jet mass prediction derived from the UDRs are propagated to the predicted \MJ\ distribution. 
The background estimation performance is first examined in the validation regions. 
Figure~\ref{fig:validation:MJ} shows the observed and predicted \MJ distributions in the validation regions,
 where in general they are seen to agree well.
The difference between the observed and predicted \MJ~distributions 
 is consistent with variations of the jet mass prediction due to correlated systematic uncertainties
 and is covered by the total uncertainty.
Figure~\ref{fig:signal:MJ} shows the predicted and observed \MJ distributions in the signal regions.
 
The statistical interpretation is based on the event yield in a signal region beyond an \MJ threshold, 
 which maximizes the sensitivity to both the \tenandsixQ.
For the 5jSR and 5jSRb\_1 signal regions, the threshold used is 0.8 TeV, 
except that for direct decay models with \mg\ < 1080 GeV, 5jSRb\_2 with \MJ\ > 0.6 TeV is found to be optimal.
For the 4jSR and 4jSRb signal regions, the \MJ\ threshold is 1.0~TeV.
The model-independent interpretation is performed in all the signal regions with the \MJ\ requirements mentioned just above.

\section{Signal systematic uncertainties}
\label{sec:syst}
The main systematic uncertainties for the predicted signal yield include the \largeR\ jet mass scale and resolution uncertainties, 
 $b$-tagging uncertainty, Monte Carlo statistical uncertainty, and luminosity uncertainty. 
The \largeR\ jet mass scale and resolution uncertainties are estimated by comparing the performance of calorimeter-based jets with the performance of track-based jets in
 data and Monte Carlo simulation samples~\cite{ATLAS-CONF-2016-035}.
The uncertainty in the predicted signal yields due to the \largeR\ jet mass scale and resolution uncertainty is as large as 24\% for signal models with \mg~= 1000 GeV, 
and decreases to 8\% for signal models with \mg~=1800 GeV.
The Monte Carlo samples reproduce the $b$-tagging efficiency measured in data with limited accuracy. 
Dedicated correction factors, derived from a comparison between \ttbar~events in data and Monte Carlo simulation,
are applied to the signal samples~\cite{PERF-2012-04}.
The uncertainty of the correction factors is propagated to a systematic uncertainty in the yields in the signal region.
This uncertainty is between 1\% and 5\% for all signal models considered in this analysis. 
Due to low acceptance, the statistical uncertainty of the signal yield predicted by the Monte Carlo samples 
 can be as large as 8\% for signal models with \mg~$\le$ 1000 GeV. 
The Monte Carlo statistical uncertainty for signal models with large \mg~is negligible.
Uncertainties in the signal acceptance due to the choices of QCD scales and PDF, and the modelling of initial-state radiation (ISR) 
 are studied. The uncertainty due to the PDF and QCD scales is found to be as large as 25\% for \mg~= 1000 GeV, 10\% for \mg~= 1700 GeV,
and a few percent for \mg~=2100 GeV. The relatively large uncertainty at \mg~= 1000 GeV is 
partly because the signal region \MJ~requirement is placed at the tail of the \MJ~distribution, which is more sensitive to scale variations.

Since signal events and background events have different kinematic distributions and jet flavour compositions,
 the presence of signal events in data can bias the predicted background yield in the signal region. 
The presence of signal events can lead to a positive contribution to the predicted background yield, which can be 
determined by studying signal Monte Carlo samples, and therefore is subtracted from the background prediction for the model-dependent interpretation.
This potential bias is not considered for the model-independent interpretation. 
As the contribution is induced by the signal events, the correction also scales with the cross section of the signal events, 
 which is equivalent to a correction of the predicted signal yield. 
The size of the correction relative to the predicted signal can be as large as 50\% for cascade decay models with 
\mx~=50~GeV, and decreases to a few percent for models with a small mass difference between the gluino and neutralino.

\section{Results}
\label{sec:result}

Table~\ref{tab:result} summarizes the predicted and observed event yields in signal regions with different \MJ requirements,
which are used to construct the likelihood function for the statistical interpretation. The number of events in each signal region's corresponding normalization region is also shown.
Modest, but not statistically significant, excesses are seen in signal regions requiring five or more jets and the 4jSR signal region.

\begin{table}[t]
\caption{ Predicted and observed yields in various search regions for a number of different \MJ~requirements. The number of events in the normalization region, $\mathrm{N}_{\mathrm{NR}}$, is also shown.}
  \begin{center}
    \begin{tabular}{ccccccccccc}
\hline
Region & $\mathrm{N}_{\mathrm{NR}}$ & $\ge~$\MJ~[TeV]  & Expected (&$\pm$& (stat.)&$\pm$& (high-\pt)&$\pm$& (low-\pt)) & Observed\\
\hline
4jSRb  & 64081 & 1.0     &  23.6 &$\pm$&  4.6 &$\pm$&  ~~6.1  &$\pm$&  1.7 & 15\\
4jSR   & 224862 & 1.0     &  ~68.2 &$\pm$&  7.6 &$\pm$&  15.8 &$\pm$&  4.4 & 82 \\
\hline
5jSRb\_1 &  2177 &  0.8     &  ~~7.0  &$\pm$&  2.4 &$\pm$&  ~~1.9  &$\pm$&  0.7 & 10\\
5jSRb\_2 & 2177 &  0.6     &  ~44.0 &$\pm$&  7.5 &$\pm$&  11.2 &$\pm$&  7.2 & 61 \\
5jSR   & 6592 & 0.8     &  18.0 &$\pm$&  3.7 &$\pm$&  ~~4.6  &$\pm$&  1.5 & 31\\
\hline

\hline
\end{tabular}
\end{center}

\label{tab:result}
\end{table}

Signal and background systematic uncertainties are incorporated as nuisance parameters. 
A frequentist procedure based on the profile likelihood ratio~\cite{Cowan:2010js} is used to evaluate the $p_{0}$-values of these excesses, 
and the results are shown in Table~\ref{tab:statistics}.
Since no significant excess is seen in any of the  signal regions,
a model-independent limit on $\sigma_{\mathrm{vis}}$,
defined as the upper limit on the number of signal events of a generic BSM model in the signal region divided by the integrated luminosity,
is calculated using a modified frequentist procedure (the {\tt   $\text{CL}_\text{s}$} method~\cite{Read:2002hq}).
The observed and expected limits are shown in Table~\ref{tab:statistics}.

Limits are set on the production of gluinos in UDD scenarios of RPV SUSY
 and are shown in Figure~\ref{fig:limit}. 
Typically, for RPV signals from the \tenQ with \mg~= 1800 GeV and 250 GeV $\le$ \mx~< 1650 GeV, 
the detector efficiency, defined as the ratio of the selection efficiency at detector level to the event-generator-level acceptance, 
is between 1.2 and 1.4, for 5jSRb with \MJ > 0.8 TeV. The detector efficiency at \mx~=1050 GeV, varies between 1.5 for \mg~= 1200 GeV to 1.2 for  \mg~= 2000 GeV.
The ratio is beyond 1 because the migration of events due to effects of resolution and efficiency at the reconstruction level. 
The search excludes a gluino with mass 1000--\lten~GeV at the 95\% confidence level (CL) in the \tenQ, 
 with the most stringent limit achieved at \mx $\gtrsim$ 1000 GeV and the weakest limit achieved at \mx $\gtrsim$ 50 GeV.
The exclusion is weaker for signal points with a small \mx or a small gap between \mx~and \mg, because these signal points have smaller jet multiplicities and hence smaller efficiencies.
For the \sixQ, the search does not exclude any specific range of gluino mass due to an upward fluctuation in the signal regions, nonetheless, the search yields a 95\% CL upper limit on the production cross section between 0.011 \ifb\ and 0.80 \ifb, in the range of 900 GeV 
 < \mx\ < 1800 GeV.

\begin{table}
\caption{
 Expected and observed limits on the signal production cross section for the signal regions.
The observed $p_{0}$-value is also shown.}

\renewcommand{\arraystretch}{1.2}
\begin{center}
\vspace{0.3cm}

\hspace*{-0.1cm} \begin{tabular}{lclclclc|c|}
\hline
Signal region & \MJ requirement & Expected limit [fb] & Observed limit [fb] & $p_{0}$-value\\
\hline
4jSRb    &  > 1.0 TeV & $0.53^{+0.20}_{-0.12}$  & 0.37 & 0.5\\
4jSR     &  > 1.0 TeV & $1.12^{+0.50}_{-0.32}$  & 1.50 & 0.24\\
5jSRb\_1 &  > 0.8 TeV & $0.24^{+0.10}_{-0.06}$  &  0.34 & 0.26\\
5jSRb\_2 &  > 0.6 TeV & $0.86^{+0.40}_{-0.20}$  &  1.32 & 0.20\\
5jSR     &  > 0.8 TeV & $0.44^{+0.18}_{-0.10}$ & 0.84 & 0.062\\
\hline
\end{tabular}
\label{tab:statistics}
\end{center}
\end{table}

\begin{figure}[!htb]
  \centering

  \subfigure[\sixQ]{
   \includegraphics[width=0.48\textwidth]{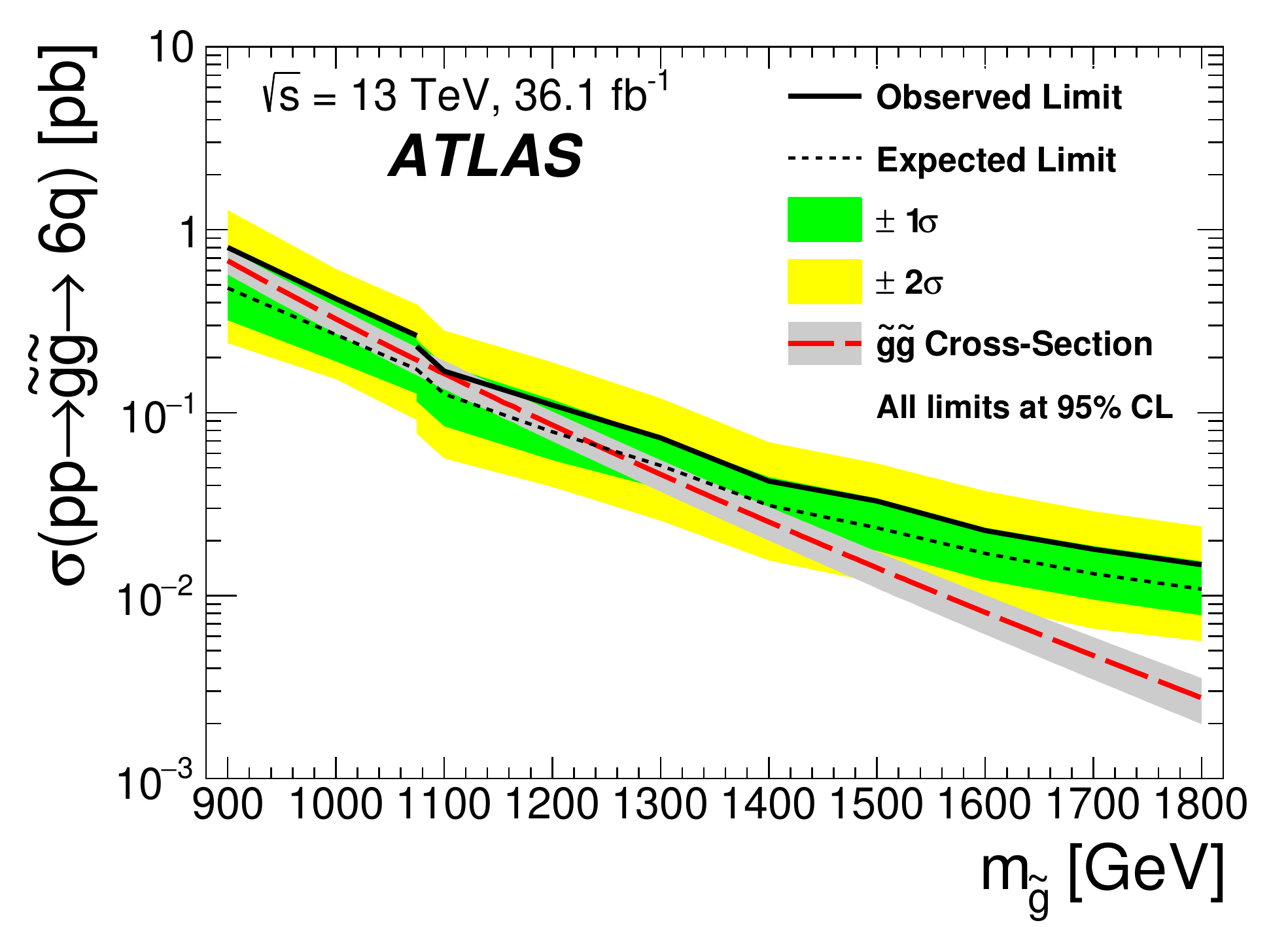}
    \label{fig:limit:6q}} 
  \subfigure[\tenQ]{
   \includegraphics[width=0.48\textwidth]{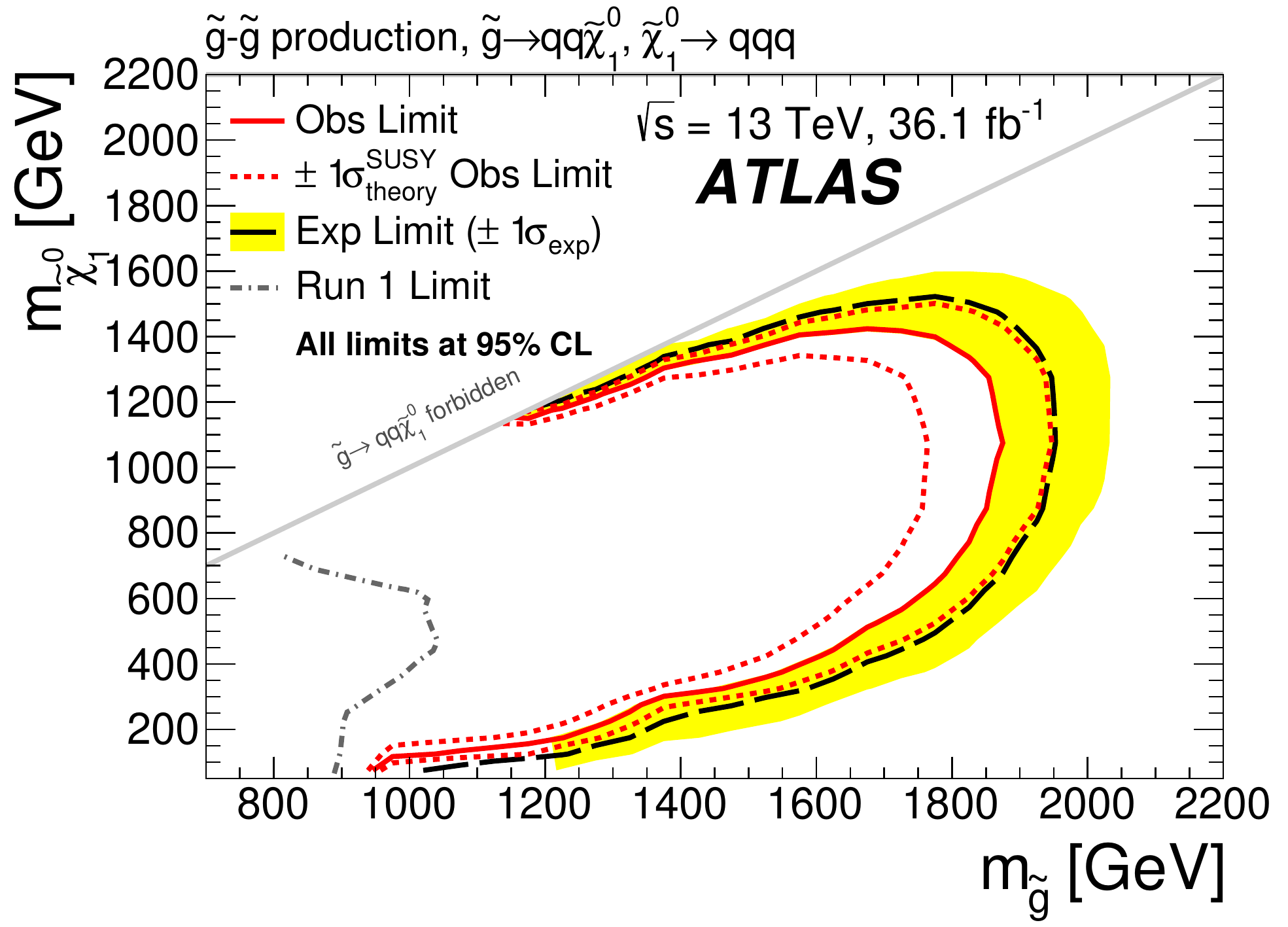}
    \label{fig:limit:10q}}\\

  \caption{
\subref{fig:limit:6q} Expected and observed cross-section limits for
the \sixQ. The discontinuities in the observed limit and~$\pm 1\sigma$ and $\pm 2\sigma$ bands
 are caused by the use of two different signal regions (5jSRb\_2 for \mg~< 1080 GeV, 5jSRb\_1 for \mg~> 1080 GeV).
The long-dashed line and the grey band surrounding it are the expected gluino pair production cross section and the associated theoretical uncertainty.
\subref{fig:limit:10q} Expected and observed exclusion contours in the (\mg,\mx)  plane for the \tenQ.
 The dashed black line shows the expected limit at 95\% CL, with
 the light (yellow) band indicating the~$\pm 1\sigma$ variations due to
 experimental uncertainties. Observed limits are indicated by red curves,
where the solid contour represents the nominal limit, and the
dotted lines are obtained by varying the signal cross section
by the renormalization and factorization scale and PDF uncertainties.
The observed limit from the Run-1 analysis~\cite{SUSY-2013-07} is also shown as a dotted-dashed line.
}
  \label{fig:limit}
\end{figure}

\FloatBarrier

\section{Conclusion}
\label{sec:conclusion}

A search for R-parity-violating SUSY signals in events with multiple jets is conducted with 36.1 \ifb of proton--proton collision data at $\sqrt{s}$ = 13 TeV collected by the ATLAS detector at the LHC. 
Distributions of events as a function of total jet mass of the four leading jets in \pt~are examined. 
No significant excess is seen in any signal region. 
Limits are set on the production of gluinos in the \tenandsixQ in the UDD scenarios of RPV SUSY.
In the \tenQ, gluinos with masses between 1000 GeV and \lten~GeV are excluded at 95\% CL, depending on the neutralino mass;
in the \sixQ, signals with a cross section of  0.011--0.8 fb\ are excluded at 95\% CL, depending on the gluino mass.
Model-independent limits are also set on the signal production cross section times branching ratio in five overlapping signal regions.
These significantly extend the limits from the 8 TeV LHC analyses.

\section*{Acknowledgements}

We thank CERN for the very successful operation of the LHC, as well as the
support staff from our institutions without whom ATLAS could not be
operated efficiently.

We acknowledge the support of ANPCyT, Argentina; YerPhI, Armenia; ARC, Australia; BMWFW and FWF, Austria; ANAS, Azerbaijan; SSTC, Belarus; CNPq and FAPESP, Brazil; NSERC, NRC and CFI, Canada; CERN; CONICYT, Chile; CAS, MOST and NSFC, China; COLCIENCIAS, Colombia; MSMT CR, MPO CR and VSC CR, Czech Republic; DNRF and DNSRC, Denmark; IN2P3-CNRS, CEA-DRF/IRFU, France; SRNSFG, Georgia; BMBF, HGF, and MPG, Germany; GSRT, Greece; RGC, Hong Kong SAR, China; ISF, I-CORE and Benoziyo Center, Israel; INFN, Italy; MEXT and JSPS, Japan; CNRST, Morocco; NWO, Netherlands; RCN, Norway; MNiSW and NCN, Poland; FCT, Portugal; MNE/IFA, Romania; MES of Russia and NRC KI, Russian Federation; JINR; MESTD, Serbia; MSSR, Slovakia; ARRS and MIZ\v{S}, Slovenia; DST/NRF, South Africa; MINECO, Spain; SRC and Wallenberg Foundation, Sweden; SERI, SNSF and Cantons of Bern and Geneva, Switzerland; MOST, Taiwan; TAEK, Turkey; STFC, United Kingdom; DOE and NSF, United States of America. In addition, individual groups and members have received support from BCKDF, the Canada Council, CANARIE, CRC, Compute Canada, FQRNT, and the Ontario Innovation Trust, Canada; EPLANET, ERC, ERDF, FP7, Horizon 2020 and Marie Sk{\l}odowska-Curie Actions, European Union; Investissements d'Avenir Labex and Idex, ANR, R{\'e}gion Auvergne and Fondation Partager le Savoir, France; DFG and AvH Foundation, Germany; Herakleitos, Thales and Aristeia programmes co-financed by EU-ESF and the Greek NSRF; BSF, GIF and Minerva, Israel; BRF, Norway; CERCA Programme Generalitat de Catalunya, Generalitat Valenciana, Spain; the Royal Society and Leverhulme Trust, United Kingdom.

The crucial computing support from all WLCG partners is acknowledged gratefully, in particular from CERN, the ATLAS Tier-1 facilities at TRIUMF (Canada), NDGF (Denmark, Norway, Sweden), CC-IN2P3 (France), KIT/GridKA (Germany), INFN-CNAF (Italy), NL-T1 (Netherlands), PIC (Spain), ASGC (Taiwan), RAL (UK) and BNL (USA), the Tier-2 facilities worldwide and large non-WLCG resource providers. Major contributors of computing resources are listed in Ref.~\cite{ATL-GEN-PUB-2016-002}.

\clearpage

\printbibliography

\FloatBarrier
\clearpage \input{atlas_authlist}
\end{document}

%% file: SUSY-2016-22-PAPER-metadata.tex
\AtlasTitle{Search for R-parity-violating supersymmetric particles in multi-jet final states produced in $p$--$p$ collisions at $\sqrt{s}~=$~13 TeV using the ATLAS detector at the LHC}

\author{The ATLAS Collaboration}

\date{\today}

\AtlasNote{ATLAS-SUSY-2016-022}

\PreprintIdNumber{CERN-EP-2017-298}

\AtlasJournalRef{Phys. Lett. B 785 (2018) 136} 
\AtlasDOI{10.1016/j.physletb.2018.08.021}

 \AtlasJournal{Phys.\ Lett.\ B.}

\AtlasAbstract{Results of a search for gluino pair production with subsequent R-parity-violating decays to quarks are presented.
This search uses 36.1 fb$^{-1}$ of data collected by the ATLAS detector in proton--proton collisions with a center-of-mass energy of $\sqrt{s}$ = 13 TeV at the LHC.
The analysis is performed using requirements on the number of jets and the number of jets tagged as containing a $b$-hadron as well as a topological observable formed by the scalar sum of masses of large-radius jets in the event.
No significant excess above the expected Standard Model background is observed.
Limits are set on the production of gluinos in models with the R-parity-violating  decays of either the gluino itself (direct decay) or the neutralino produced in the R-parity-conserving gluino decay (cascade decay).
In the \tenQ, gluino masses below 1850 GeV are excluded for 1000 GeV neutralino mass.
For the \sixQ, the 95\% confidence level upper limit on the   cross section times branching ratio
 varies between 0.80 fb~at \mg~= 900 GeV and 0.011 fb~at \mg~= 1800 GeV.}

%% file: atlas_authlist.tex
 
\begin{flushleft}
{\Large The ATLAS Collaboration}

\bigskip

M.~Aaboud$^\textrm{\scriptsize 34d}$,    
G.~Aad$^\textrm{\scriptsize 99}$,    
B.~Abbott$^\textrm{\scriptsize 124}$,    
O.~Abdinov$^\textrm{\scriptsize 13,*}$,    
B.~Abeloos$^\textrm{\scriptsize 128}$,    
S.H.~Abidi$^\textrm{\scriptsize 165}$,    
O.S.~AbouZeid$^\textrm{\scriptsize 143}$,    
N.L.~Abraham$^\textrm{\scriptsize 153}$,    
H.~Abramowicz$^\textrm{\scriptsize 159}$,    
H.~Abreu$^\textrm{\scriptsize 158}$
% ...
\bigskip
\\

% ...
$^{13}$Institute of Physics, Azerbaijan Academy of Sciences, Baku; Azerbaijan.\\
% ...
$^{34}$$^{(a)}$Facult\'e des Sciences Ain Chock, R\'eseau Universitaire de Physique des Hautes Energies - Universit\'e Hassan II, Casablanca;$^{(b)}$Centre National de l'Energie des Sciences Techniques Nucleaires (CNESTEN), Rabat;$^{(c)}$Facult\'e des Sciences Semlalia, Universit\'e Cadi Ayyad, LPHEA-Marrakech;$^{(d)}$Facult\'e des Sciences, Universit\'e Mohamed Premier and LPTPM, Oujda;$^{(e)}$Facult\'e des sciences, Universit\'e Mohammed V, Rabat; Morocco.\\
% ...
$^{99}$CPPM, Aix-Marseille Universit\'e, CNRS/IN2P3, Marseille; France.\\
% ...
$^{124}$Homer L. Dodge Department of Physics and Astronomy, University of Oklahoma, Norman OK; United States of America.\\
% ...
$^{128}$LAL, Universit\'e Paris-Sud, CNRS/IN2P3, Universit\'e Paris-Saclay, Orsay; France.\\
% ...
$^{143}$Santa Cruz Institute for Particle Physics, University of California Santa Cruz, Santa Cruz CA; United States of America.\\
% ...
$^{153}$Department of Physics and Astronomy, University of Sussex, Brighton; United Kingdom.\\
% ...
$^{158}$Department of Physics, Technion, Israel Institute of Technology, Haifa; Israel.\\
$^{159}$Raymond and Beverly Sackler School of Physics and Astronomy, Tel Aviv University, Tel Aviv; Israel.\\
% ...
$^{165}$Department of Physics, University of Toronto, Toronto ON; Canada.\\
% ...
$^{a}$ Also at  Department of Physics, University of Malaya, Kuala Lumpur; Malaysia.\\
$^{b}$ Also at Borough of Manhattan Community College, City University of New York, NY; United States of America.\\
$^{c}$ Also at Centre for High Performance Computing, CSIR Campus, Rosebank, Cape Town; South Africa.\\
$^{d}$ Also at CERN, Geneva; Switzerland.\\
$^{e}$ Also at CPPM, Aix-Marseille Universit\'e, CNRS/IN2P3, Marseille; France.\\
% ...
$^{*}$ Deceased

\end{flushleft}
